\documentclass{aa}
\usepackage{txfonts}

\usepackage{graphicx}
\usepackage{natbib}

\bibpunct{(}{)}{;}{a}{}{,}

\usepackage{color}

\newcommand{\red}[1]  {\textcolor{red}{#1}}
\newcommand{\blue}[1] {\textcolor{blue}{#1}}

\newcommand{\ie} {{\em i.e.}}
\newcommand{\eg} {{\em e.g.}}

\newcommand{\HHth}{HH\,30}
\newcommand{\emm}[1]{\ensuremath{#1}}   
\newcommand{\emr}[1]{\emm{\mathrm{#1}}} 
\newcommand{\about}{\emm{\sim}}
\newcommand{\unit}[1]{\emm{\, \emr{#1}}}
\newcommand{\nds}[1]{\emm{\displaystyle#1}} 
\newcommand{\paren}[1]  {\nds{\left(  #1 \right) }} 
\newcommand{\cbrace}[1]{\nds{\left\{ #1 \right\}}} 
\newcommand{\hbrace}[1]{\nds{\left\{ #1 \right.}}  
\newcommand{\abs}[1]{\emm{\left| #1 \right|}} 
\newcommand{\ala}{\mbox{\rlap{\hbox{\lower4pt\hbox{$\sim$}}}\hbox{$<$}}} 
\newcommand{\pc}{\unit{pc}}
\newcommand{\au}{\unit{AU}}
\newcommand{\nm}{\unit{nm}}
\newcommand{\mm}{\unit{mm}}
\newcommand{\K}{\unit{K}}
\newcommand{\kms}{\unit{km\,s^{-1}}}
\newcommand{\Jykms}{\unit{Jy\,km\,s^{-1}}}

\newcommand{\Lsun}{\unit{L_\odot}}
\newcommand{\Msun}{\unit{M_\odot}}
\newcommand{\Msunpyr} {\unit{M_\odot\,yr^{-1}}}
\renewcommand{\deg}{\emm{\degr}}

\newcommand{\pccm}{\unit{cm^{-3}}}
\newcommand{\emiss}{\unit{cm^2\,g^{-1}}}
\newcommand{\HH}  {\mbox{$\rm H_2$}}     
\newcommand{\thCO}{\mbox{$\rm ^{13}CO$}} 
\newcommand{\twCO}{\mbox{$\rm ^{12}CO$}} 
\newcommand{\CeiO}{\mbox{$\rm C^{18}O$}} 
\newcommand{\HCOp}{\mbox{$\rm HCO^{+}$}} 
\newcommand{\Jone}{\mbox{J=1--0}}
\newcommand{\Jtwo}{\mbox{J=2--1}}

\newcommand{\nht}{\ifmmode {{\rm NH}_3} \else {NH{\bas 3}} \fi}

\newcommand{\as}{\ifmmode {^{\scriptscriptstyle\prime\prime}}
        \else $^{\scriptscriptstyle\prime\prime}$\fi}

\newcommand{\TableObs}{%
  \begin{table*}
    \caption{Observation parameters.}
    \begin{center}
      \begin{tabular}{lll}
        \hline
        Phase center & $\alpha_{2000} = 04^h31^m37.47^s $ & $ \delta_{2000} = 18\degr 12' 24.2''$\\
        \hline
      \end{tabular}
      \medskip{}
      \begin{tabular}{rrcrccccr}
        \hline
        Molecule \& Line & Frequency  &   Beam     & PA     & $\delta v^{a}$ & \multicolumn{2}{c}{1$\sigma$ noise levels$^{a}$} & Int. Time$^{b}$ & \multicolumn{1}{c}{Obs. Date} \\
                         & GHz        & arcsec$^2$ & $\deg$ &        \kms{} &      mJy/Beam        &        K        &           hours & \\
        \hline
        \HCOp{} \Jone{}  &  89.188 & $4.28 \times 3.33$ & 48 & 0.4 & 10 & 0.11 & ~6 & 1997--1998 \\
        \twCO{} \Jtwo{}  & 230.538 & $1.78 \times 1.16$ & 14 & 0.4 & 29 & 0.32 & ~6 & 1997--1998 \\
        \CeiO{} \Jtwo{}  & 109.782 & $3.49 \times 2.82$ & 35 & 0.3 & ~9 & 0.09 & 14 & 2000--2002 \\
        \thCO{} \Jone{}  & 110.201 & $3.48 \times 2.81$ & 34 & 0.3 & ~9 & 0.09 & 14 & 2000--2002 \\
        \thCO{} \Jtwo{}  & 220.399 & $1.80 \times 1.56$ & 26 & 0.3 & 22 & 0.20 & 14 & 2000--2002 \\
        \hline
      \end{tabular}
    \end{center}
    \emph{a} The rms noises are computed from the channel maps featured
    in Fig.~\ref{fig:13co-c18o} and~\ref{fig:12co21}, \ie{} using natural
    weighting and the $\delta v$ channel resolutions of the maps. The
    original correlator resolution are between 3 and 6 times higher than the one shown here. \\
    \emph{b} This is the total \emph{on--source} observing time of useful data as if all
    observations were done with 5 antennas.
    \label{tab:obs}
  \end{table*}}

\newcommand{\TableContinuum}{%
  \begin{table*}
    \centering %
    \caption{Continuum fluxes and sizes as a function of the frequency.}
    \begin{tabular}{rrrcccc}
      \hline
      Frequency & Wavelengths & Flux  & Gaussian FWHM & Gaussian PA & Beam FWHM & Beam PA \\
      (GHz)     & (mm)        & (mJy) & ($''$)        & ($\deg$)    & ($''$)    & ($\deg$)\\
      \hline
         89.2 &      3.40 &  $2.5 \pm 0.5$ & $< 2$ (\ie{} unresolved)             & ---        & $4.28 \times 3.33$ & 47 \\
        110.0 &      2.65 &  $3.8 \pm 0.2$ & $< 2$ (\ie{} unresolved)             & ---        & $2.86 \times 2.25$ & 36 \\
        220.4 &      1.35 & $18.1 \pm 0.7$ & $1.28 \pm 0.10 \times 0.53 \pm 0.12$ & $-51\pm~6$ & $1.82 \times 1.16$ & 16 \\
        230.5 &      1.30 & $23.0 \pm 1.4$ & $1.10 \pm 0.20 \times 0.62 \pm 0.12$ & $-64\pm13$ & $1.42 \times 1.22$ & 30 \\
  220.4+230.5 & 1.35+1.30 & $17.3 \pm 0.6$ & $1.26 \pm 0.11 \times 0.58 \pm 0.09$ & $-50\pm~5$ & $1.29 \times 0.90$ & 33 \\
      \hline
    \end{tabular}
    \label{tab:cont}
  \end{table*}}

\newcommand{\law}[1]{\multicolumn{4}{|c|}{\rule[-0.5cm]{0pt}{1.4cm} \blue{\protect#1}} & & & }
\newcommand{\lawbis}[1]{\multicolumn{6}{|c|}{\rule[-0.5cm]{0pt}{1.4cm} \blue{\protect#1}}}
\newcommand{\lawter}[1]{\multicolumn{4}{|c|}{\rule[-0.5cm]{0pt}{1.4cm} \blue{\protect#1}}}

\newcommand{\TableDiskPropertiesComparison}{%
  \begin{table*}[f]
    \centering %
    \caption{Comparison of mm and optical/NIR properties. The upper part of
      the table shows parameters whose value is robustly deduced from the
      millimeter data (geometry, velocity law and stellar mass) while the
      bottom part displays parameters less well constrained due to the
      limited vertical resolution of this edge-on disk. Most parameters are
      deduced from the analysis of the \thCO{} emission, except for the
      parameters of the \HH{} surface density law deduced from the
      continuum emission.}
    \begin{tabular}{|l|rrl|c|c|c|}
      \hline
      & \multicolumn{3}{c|}{This work (\red{\thCO{} \& continuum data})}
      & \parbox{2.5cm}{\centering{} \citet{burrows96} \red{HST}}
      & \parbox{2.0cm}{\centering{} \citet{cotera01} \red{HST}}
      & \parbox{2.0cm}{\centering{} \citet{wood02} \red{HST+SED}} \\
      \hline
      Assumed Distance & D (pc)~= & 140 & & 140 & 140 & 140 \\
      \hline
      Systemic velocity        & $V_\mathrm{lsr}$ (km.s$^{-1}$)~=  & 7.25 & $\pm 0.04$ & --- & --- & --- \\
      Orientation (disk axis)  & PA~($^{\circ}$)  = & 32 & $\pm 2$ &   32 & 32 & 32 \\
      Inclination (disk plane) & $i$~($^{\circ}$) = & \parbox{0.5cm}{\centering{} \hbrace{\parbox{0.3cm}{81\\ 84}}} & \parbox{2.25cm}{$\pm 3$ (Best fit) \\ $\pm 3$ (Canonical fit)} & 82.5 & 84 (assumed)& 84 (assumed) \\
      \hline
      Outer radius             & \red{$R_\mathrm{out}$ (AU)~=} & \red{420} & \red{$\pm 25$} & \parbox{3cm}{\centering{} \hbrace{\parbox{2.75cm}{250 (Canonical fit) \\ 425 (Best fit)}}} & 200 (assumed) & 200 (assumed) \\
      Turbulent linewidth      & $\Delta v$ (km.s$^{-1}$)~=    & 0.23& $\pm 0.03$  & --- & --- & --- \\
      \hline
      \law{Velocity law:~~~~~~$V(r) = V_{100} \paren{\frac{r}{100\,\rm{AU}}}^{-v}$}\\
      Velocity at 100 AU     & $V_{100}$ (km.s$^{-1}$)~= &   $ 2.00$  &      $\pm 0.09$  &  --- & --- & --- \\
      Velocity exponent      & $v$~=                     &      0.50  &      $\pm 0.06$  &  --- & --- & --- \\
      Stellar mass           & \red{M$_*$ ($\Msun$)~=}   & \red{0.45} & \red{$\pm 0.04$} & 0.67 (assumed) & 0.5 (assumed) & 0.5 (assumed) \\
      \hline
      \hline
      \law{Scale Height law:~~$H(r) = H_{100} \paren{\frac{r}{100\,\rm{AU}}}^{-h}$}\\
      Scale Height at 100 AU & $H_{100}$~(AU) &  22 & & 22 ($= 15.5\sqrt{2}$) & 21 ($= 15\sqrt{2}$) & 24 ($= 17\sqrt{2}$) \\
      Scale Height exponent  & $h$            & 1.25 & (assumed)  &                1.45 &    1.29 ($= 58/45$) & 1.25 (assumed) \\
      \hline
      \law{Temperature law:~~~$T(r)~= T_{100} \paren{\frac{r}{100\,\rm{AU}}}^{-q}$} \\
      Temperature at 100 AU  & $T_{100}$ (K)~= & 12 & $\pm 1$ & 34 (Disk surface) & --- & --- \\
      Temperature exponent   & $q~\simeq$ & 0.55 & $\pm 0.07$ & 0.1 ($=3-2h$) & 0.4 ($=3-2h$) & 0.5 ($=3-2h$) \\
      \hline %
      \law{\HH{} Surface Density law:~~$\Sigma(r) = \Sigma_{100} \paren{\frac{r}{100\,\rm{AU}}}^{- p}$}\\
      Surf. dens. at 100 AU  & $\Sigma_{100}$~(cm$^{-2}$)~= & $3.6$ & $\pm 0.6~10^{22}$ & $5.5~10^{21}$ & --- & --- \\
                             & $\Sigma_{100}$~(g.cm$^{-2}$)~= & 0.16 & $\pm 0.03$ & 0.024 & --- & --- \\
      Surf. dens. exponent   & $p$~$\simeq$  & 1.0 & & 0.75 ($=s-h$) & 1.08 ($=s-h$) & 1 ($=s-h$) \\
      Density exponent       & $s~\simeq$  &  2.2 &  & 2.2 & 2.37 (assumed) & 2.25 (assumed) \\
      \hline
    \end{tabular}
    \label{tab:mm-nir-comp}
  \end{table*}}

\newcommand{\TableDustProperties}{%
  \begin{table*}[f]
    \centering %
    \caption{\HHth{} mm dust properties. The second column displays the
      best fit results while the third column shows the canonical fit
      results (\eg{} obtained with a fixed outer radius of 420\au{}, see
      text for details). The total mass is calculated assuming a 
      gas-to-dust ratio of 100 and using the kinetic temperature derived 
      from the $^{13}$CO analysis ($T_{100} = 12$ K, $q = 0.55$).}
    \begin{tabular}{|l|rrl|rl|}
      \hline
      \lawbis{Dust:~~$\kappa_\nu = \kappa_o\times(\frac{\nu}{10^{12}\,\emr{Hz}})^{\beta}$} \\
      Absorption law        & $\kappa_o$~(cm$^{2}$.g$^{-1}$)$~=$         & $0.1$       & (assumed) & & \\
      Dust exponent         & $\beta$~=                                  & $ 0.4 $     & $\pm 0.1$ & $0.5$ & $\pm 0.1$ \\
      Absorption at 230~GHz & $\kappa(230)$~(cm$^{2}$.g$^{-1}$)$~=$      & \multicolumn{2}{c|}{0.055} &  \multicolumn{2}{c|}{0.048} \\
      Dust disk size        & $R_d$ (AU)~=                       &  145           & $\pm 20$           & 420            & (assumed) \\
      Surface Density (H$_2$) &  $\Sigma_{100}$ (cm$^{-2}$)~=    & $8.6\,10^{22}$ & $\pm 1.4\,10^{22}$ & $3.6\,10^{22}$ & $\pm 0.6\,10^{22}$ \\
      Exponent                & $p$~=                             & 0              & $\pm 0.5$          &  1.0           & $\pm 0.1$ \\
      Total mass       & \red{$M_\emr{disk}$ ($\Msun$)~$\simeq$} & $2.7\,10^{-3}$ & $\pm 0.4\,10^{-3}$ & \red{$4.8\,10^{-3}$} & \red{$\pm 0.8\,10^{-3}$} \\
      \hline
    \end{tabular}
    \label{tab:mm-dust}
  \end{table*}}

\newcommand{\TableOutflowProperties}{%
  \begin{table}
    \centering %
    \caption{Outflow properties. The values in this table are model
      dependent. Furthermore they do not results from a fit. They thus are very
      different in nature from the values in the tables describing
      the disk which results from a validated fitting method.}
    \tiny{%
      \begin{tabular}{|l|rcc|}
        \hline
        & & ``Best'' & Possible range \\
        \hline
        Assumed Distance & D (pc)~= & 140 & (assumed) \\
        \hline
        Systemic velocity    & $V_\emr{lsr}$ (km.s$^{-1}$)~=  & 7.25 & $\pm 0.04$ \\
        Orientation          & PA~(\deg{})  =   & 32 & $\pm 2$ \\
        Inclination          & $i$~(\deg{}) =   & -1 & $\pm 1$ \\
        Half--opening angle  & $\theta_\emr{max}$~(\deg{}) = & 30 & $\pm 2$\\
        \hline
        Turbulent linewidth  & $\Delta v$ (\kms{})~=    & 0.40 & $\pm 0.15$ \\
        \hline
        \lawter{Velocity law:~~~~~~$V_r = V_\emr{rad}, V_\theta = 0, V_\phi = V_\emr{rot_{200}}\,r^v$} \\
        Radial velocity  & $V_\emr{rad}$ (\kms{})~=    & 11.5 & $\pm 0.5$ \\
        Rotation velocity  at 200\au{} & $V_\emr{rot_{200}}$ (\kms{})~=    &  0 & $< 1$ \\
        \hline
        \lawter{Volume density law:~~~~~~$n \propto \exp\cbrace{-\paren{\frac{r-z*\tan(\theta_\emr{max})}{w_0}}^2} \exp\cbrace{-\paren{\frac{z}{H_0}}^2}$} \\
        Width & $w_0$ $('')$~= & 0.3 & $\le 0.3$ \\
        Scale Height & $H_0$ $('')$~= & 3 & $\pm 0.5$ \\
        \hline
      \end{tabular}}
    \label{tab:outflow}
  \end{table}}

\newcommand{\TableJetOutflow}{%
  \begin{table}
    \caption{\HHth{} jet and outflow parameters in the northern lobe,
      measured in a region of $\sim 6''$ or $\sim 800\au{}$ from the central source (see
      text).}
    \begin{tabular}{lcc}
      \hline
      Quantity                      & Jet               & Outflow   \\
      \hline
      Mass (\Msun)                  & $2\,10^{-8}$      & $2\,10^{-5}$ \\
      Velocity (\kms)               & 200               & 12 \\
      \\
      Dynamical timescale (yr)      & 20                & 320 \\
      Mass flux (\Msunpyr)          & $1.0\,10^{-9}$    & $6.3\,10^{-8}$ \\
      Momentum (\Msun\kms)          & $4.0\,10^{-6}$    & $2.4\,10^{-4}$ \\
      Momentum flux (\Msun\kms/yr)  & $2.6\,10^{-7}$    & $7.5\,10^{-7}$ \\
      \hline
    \end{tabular}\\
    \label{tab:jet-flow}
  \end{table}}

\newcommand{\FigObsSum}{%
  \begin{figure*}
    \centering%
    \includegraphics[height=\hsize,angle=90]{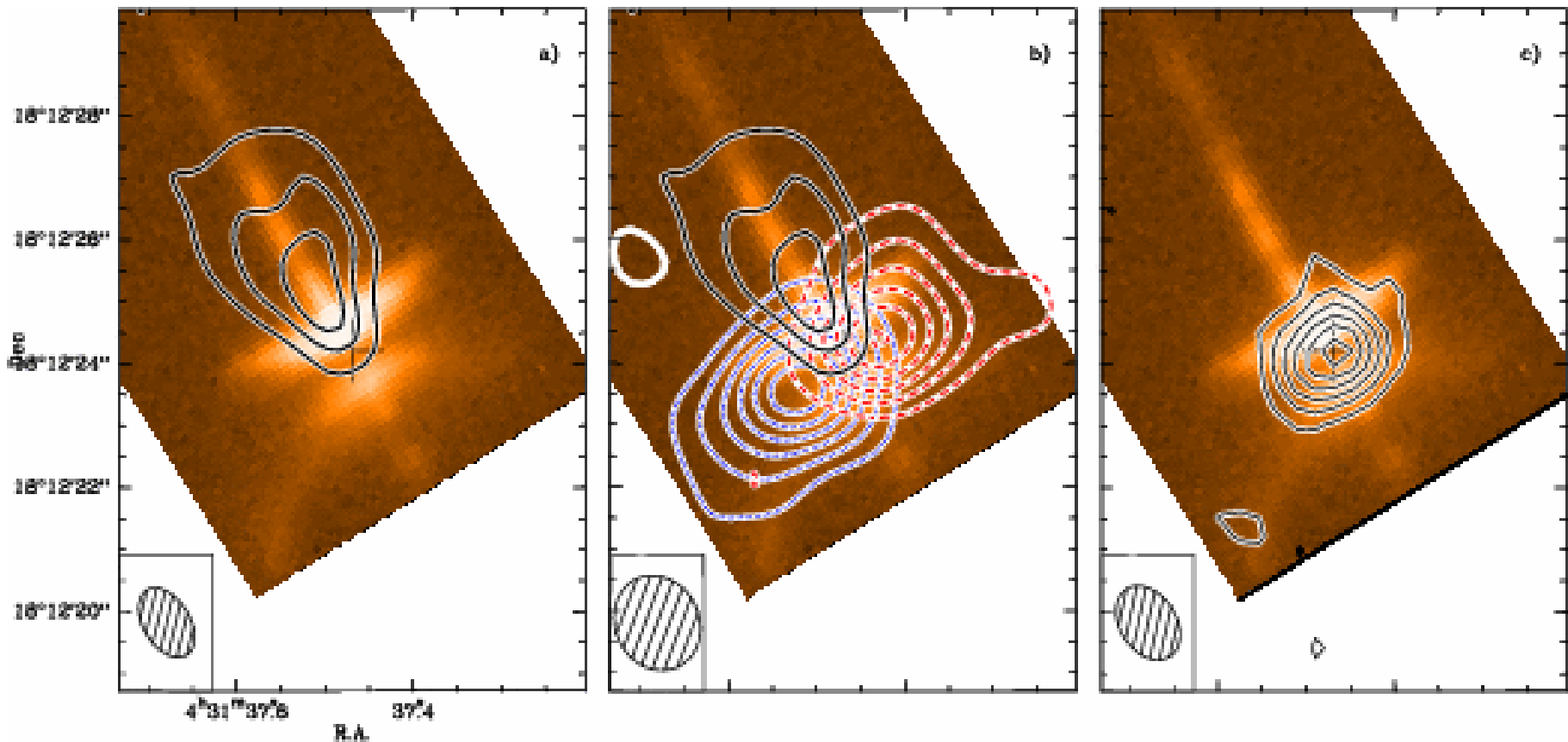}
    \caption{Left panel: Contours of the \twCO{} \Jtwo{} emission of \HHth{}
      are plotted over a composite HST image~\citep[670 and 787.7\nm{},
      see][]{burrows96}. Emission is only integrated at extreme velocities
      ($\leq 4$\kms{} and $ \geq 11$\kms) to avoid contamination by the
      disk emission.  Contours are at the 3-sigma level (48~mJy/Beam) and
      the spatial resolution is $1.23 \times 0.75''$ PA $31\deg$ (compared
      to a pixel size of $0.1''$ for the HST image).  Medium panel :
      Superimposition of 3-sigma level contours 1) of the outflow emission
      as traced by the $^{12}$CO(2--1) emission integrated at extreme
      velocities (cf.\ left panel) and 2) of the blue--shifted from 4.6 to
      7.2\kms{} (blue dotted line, 29~mJy/Beam) and red--shifted from 7.2
      to 9.8\kms{} (red dashed line, 29~mJy/Beam) emission of the disk as
      traced by the $^{13}$CO(2--1) line. The spatial resolution of the
      \thCO{}~map is $1.57 \times 1.35''$ PA $25 \deg$. Right panel:
      Superimposition of the 3-sigma level (1.4~mJy/Beam) of the merged
      1.30\,mm and 1.35\,mm{} continuum emission over the same composite
      HST image. The spatial resolution is $1.29 \times 0.91''$ PA $33
      \deg$.}
    \label{fig:pdbi-on-hst}
  \end{figure*}}

\newcommand{\FigIsotCO}{%
  \begin{figure*}
    \centering%
    \includegraphics[height=14.5cm,angle=270]{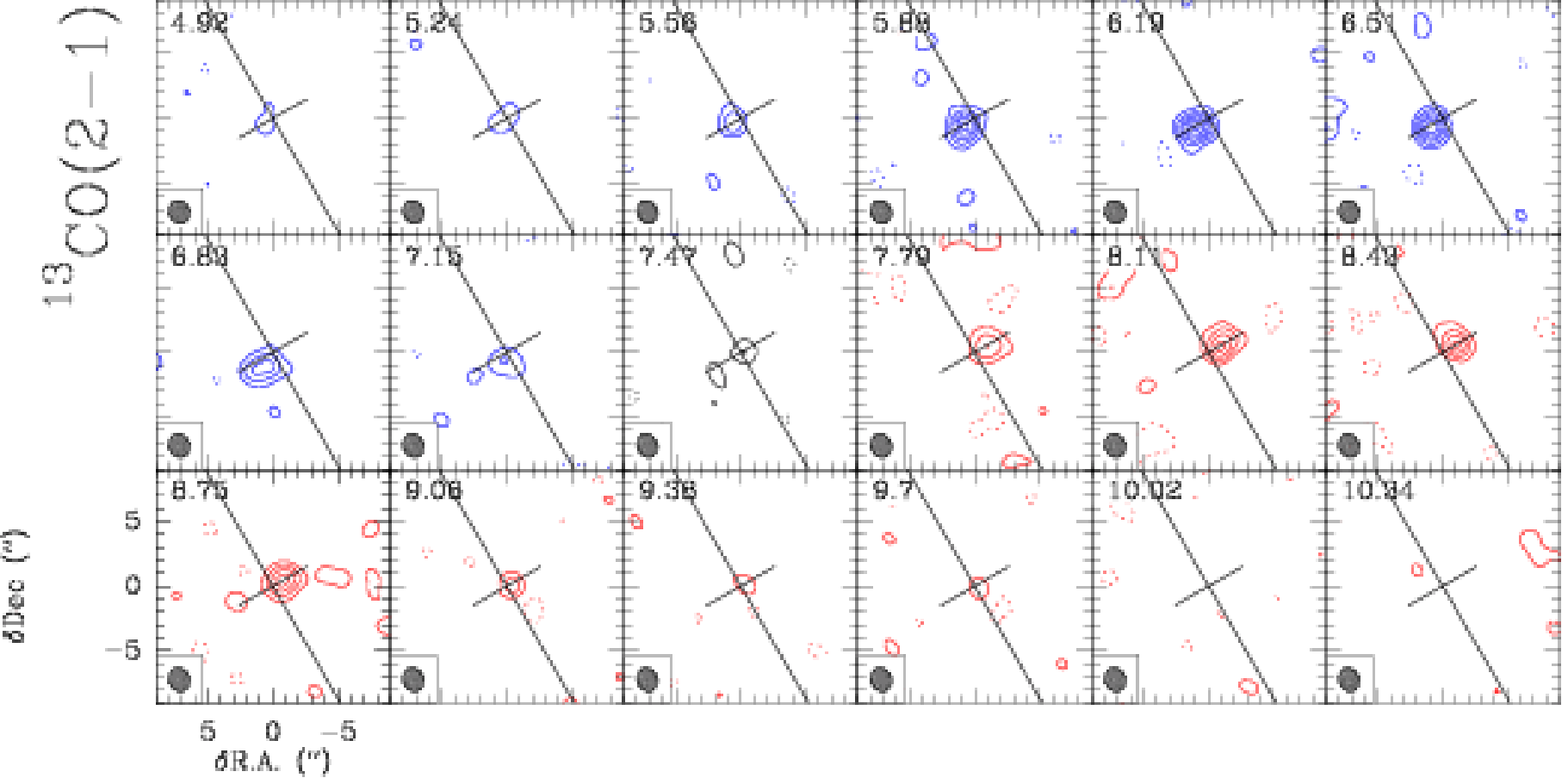}
    \includegraphics[height=14.5cm,angle=270]{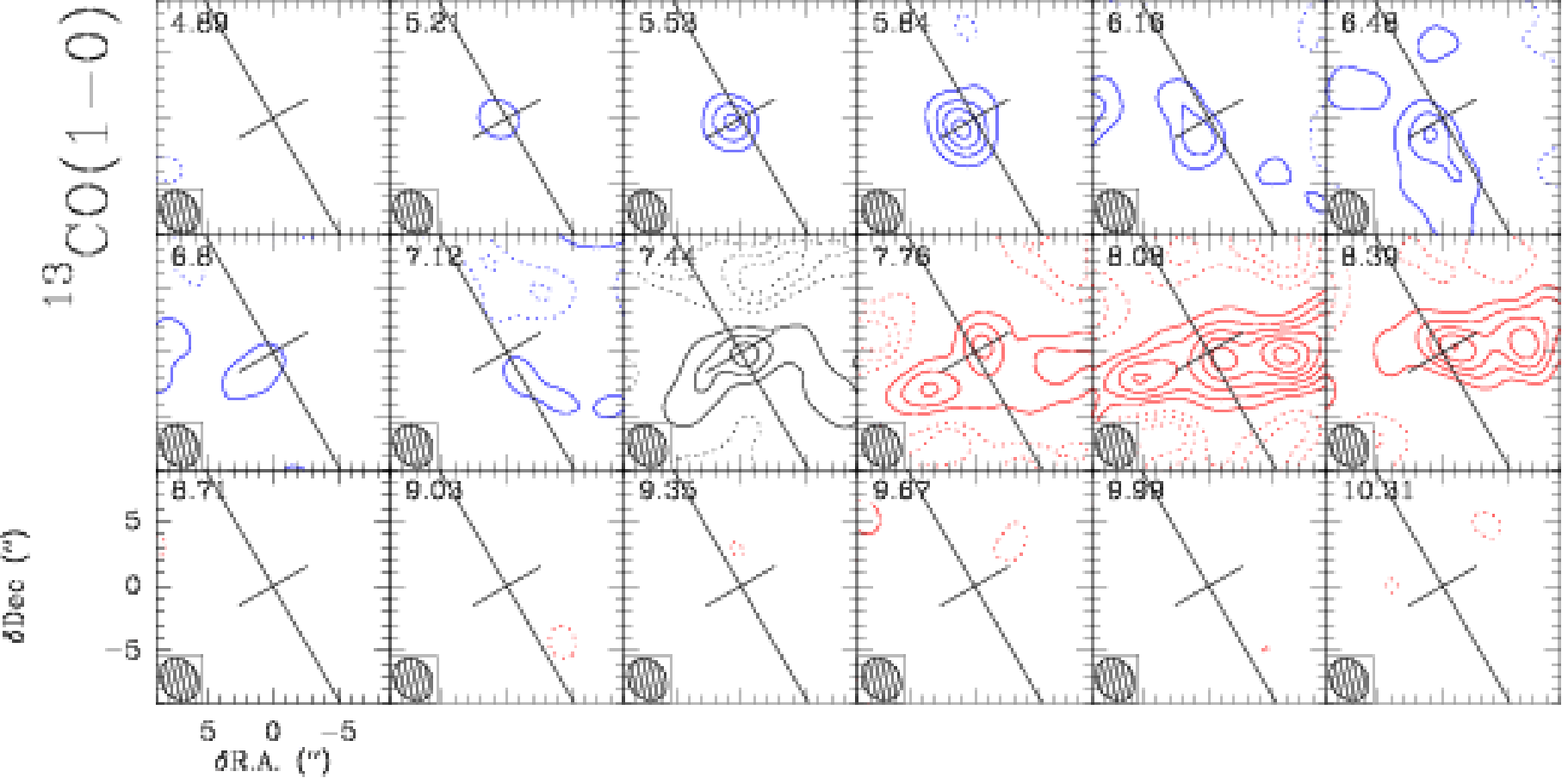}
    \includegraphics[height=14.5cm,angle=270]{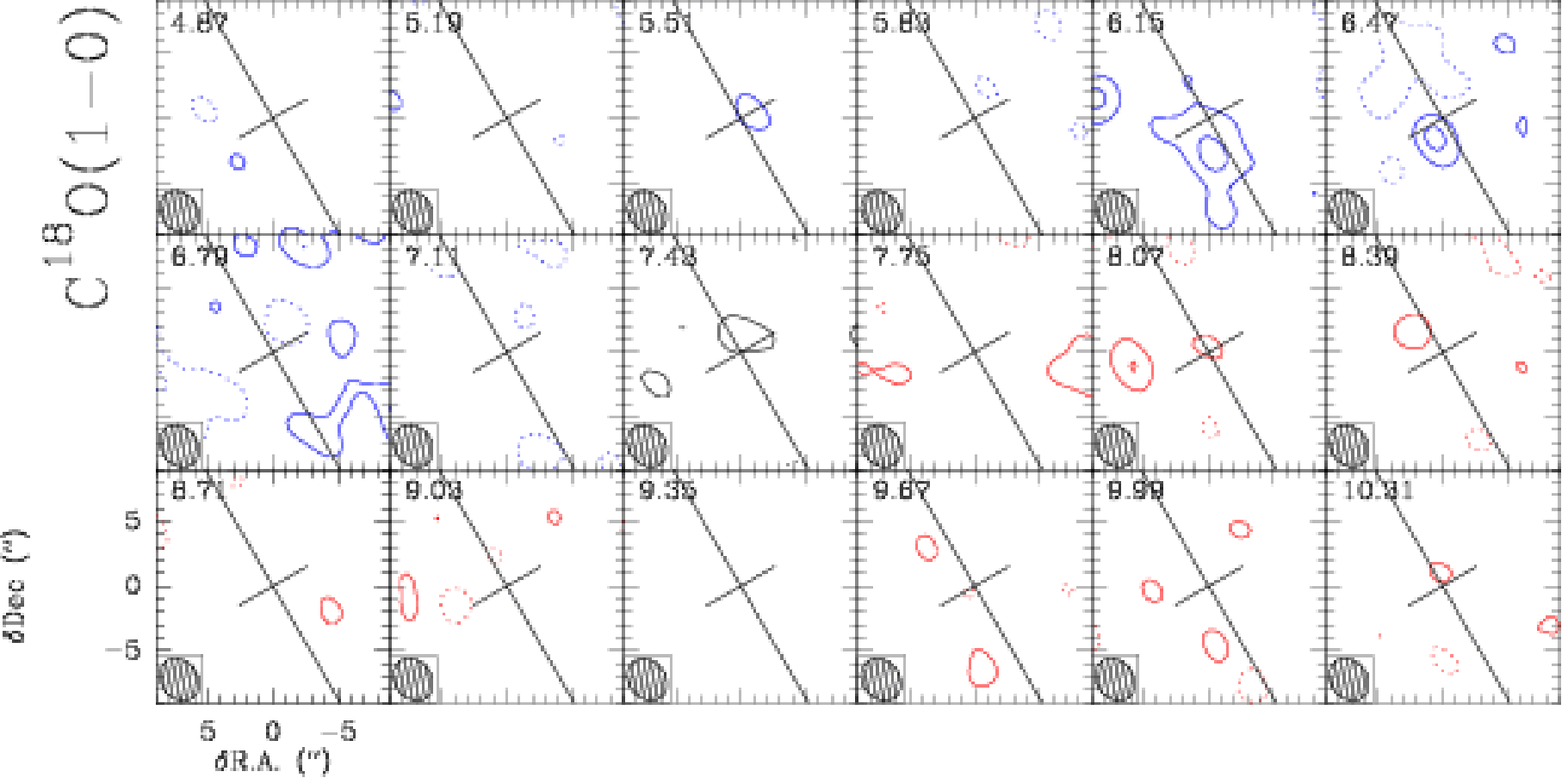}
    \caption{Line emission of the CO isotopologues in \HHth{}. From top to
      bottom: \thCO{}~\Jtwo{}, \thCO{}~\Jone{}, and \CeiO{}~\Jone{}
      emissions. The channel maps are centered near 7.25\kms{}, the \HHth{}
      systemic velocity. Blue and red contours respectively indicates the
      blue and red--shifted channels. Channel width is 0.32\kms{}. Plain and
      dotted lines respectively show positive and negative contours.
      Contour spacing corresponds to 3~$\sigma$ for \thCO{}~\Jone{}
      and~\Jtwo{} and 2~$\sigma$ for \CeiO{}~\Jone{}. Noise levels and
      spatial resolutions may be found in Table~\ref{tab:obs}. The cross
      indicates the position and orientation 1) of the continuum emission
      at 1.3\mm{} and, 2) of the optical jet.}
    \label{fig:13co-c18o}
  \end{figure*}}

\newcommand{\FigMainCO}{%
  \begin{figure*}
    \centering %
    \includegraphics[height=18cm,angle=270]{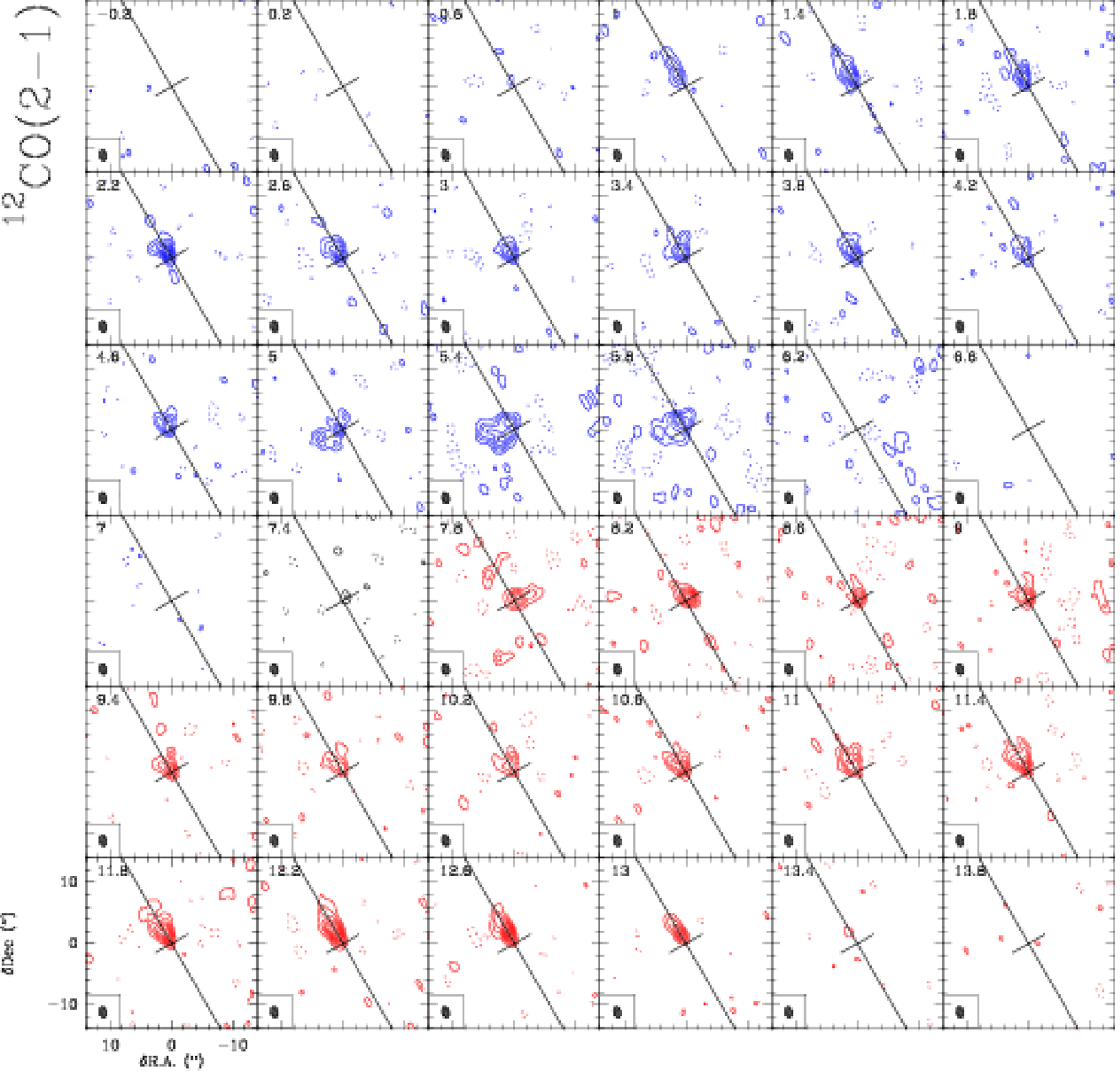}
    \caption{Same as Fig.~\ref{fig:13co-c18o} for the emission of the
      \twCO{}~\Jtwo{} line. Spectral resolution is 0.4\kms{} and contour
      spacing correspond to 3~$\sigma$ (cf.\ Table~\ref{tab:obs} for more
      details).}
      \label{fig:12co21}
  \end{figure*}}

\newcommand{\FigHCOp}{%
  \begin{figure}
    \centering %
    \includegraphics[height=\hsize{},angle=270]{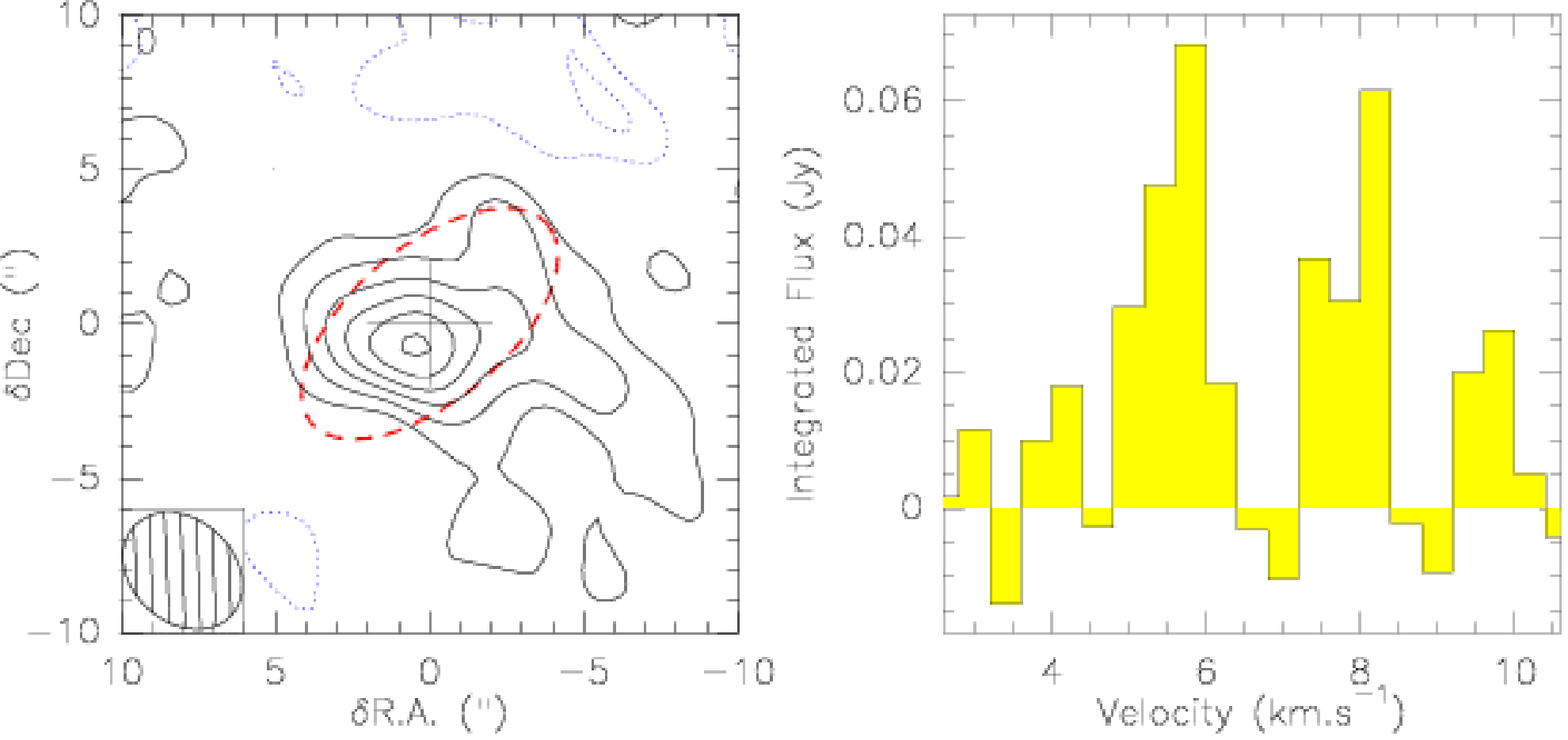}
    \caption{\textbf{Left:} Integrated emission of the \HCOp{}~\Jone{} line
      in the 5 to 8\kms{} velocity interval. Contour spacing corresponds to
      1~$\sigma$ (15~mJy/beam\kms{}). The cross indicates the position of
      the continuum emission at 1.3\mm{}. \textbf{Right:} \HCOp{}~\Jone{}
      spectrum of the flux integrated over the ellipse shown as a red dash
      contour on the left panel (size: $10''\times 5''$, with a major axis
      at PA: $-50\deg$).}
      \label{fig:hcop10}
  \end{figure}}

\newcommand{\FigDiskPV}{%
  \begin{figure}
    \centering %
    \includegraphics[height=0.45\hsize,angle=270]{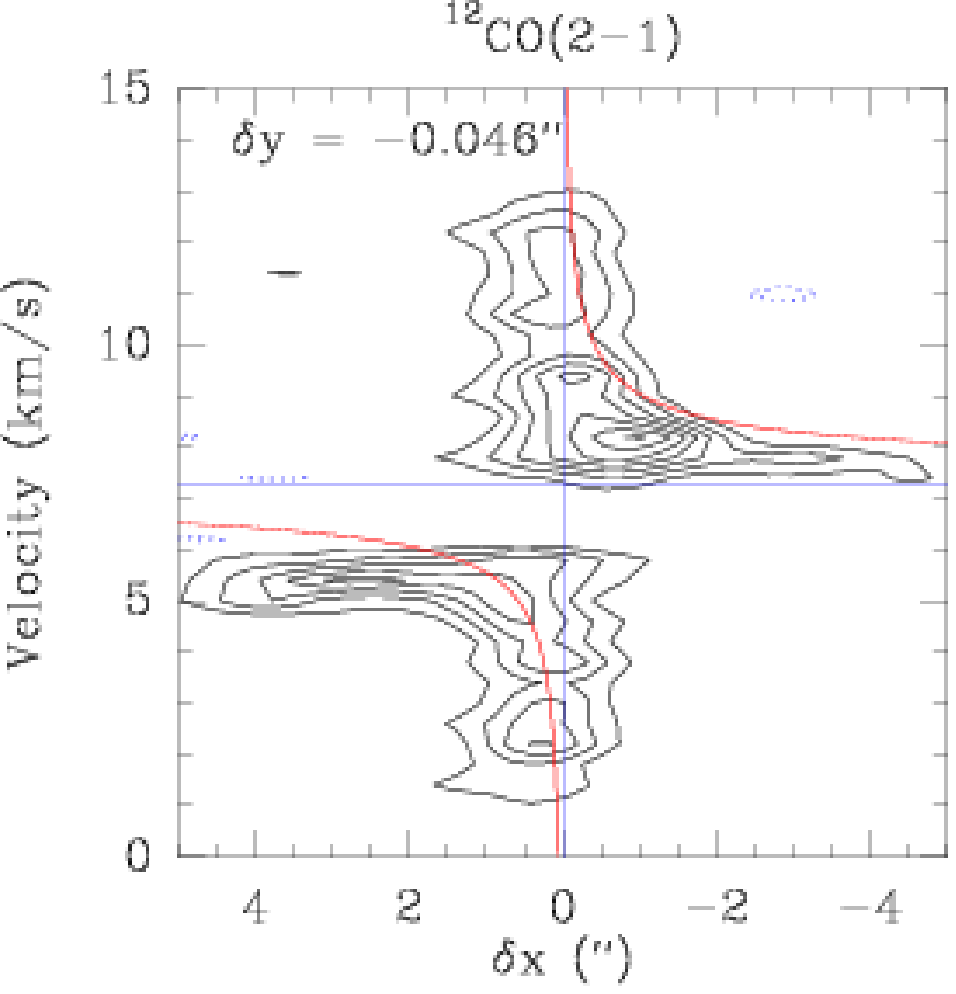}
    \hfill{}
    \includegraphics[height=0.45\hsize,angle=270]{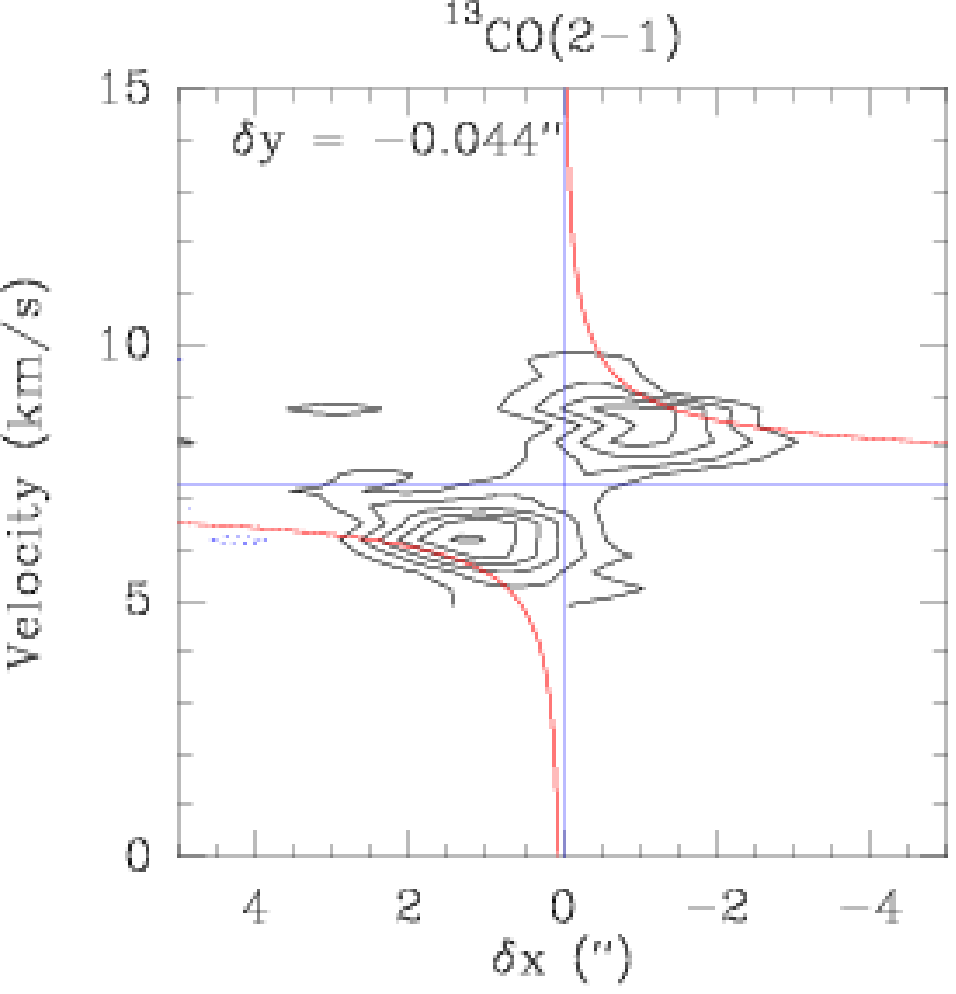}
    \caption{Position--velocity diagrams along
      the disk axis of the \twCO{}~\Jtwo{} (left panel, contour spacings
      set to 90 mJy/Beam or \about{} 3$\sigma$) and \thCO{}~\Jtwo{} (right
      panel, contour spacings set to 60 mJy/Beam or \about{} 3$\sigma$)
      emissions. The blue horizontal and vertical lines respectively
      indicate the systemic velocity and the center of the disk as seen in
      the continuum. The red curves show the theoretical Keplerian velocity
      for a $0.45~\Msun$ star.}
    \label{fig:disk-pv}
  \end{figure}}

\newcommand{\FigMainCOFlux}{%
  \begin{figure*}
    \centering %
    \includegraphics[height=0.325\hsize{},angle=270]{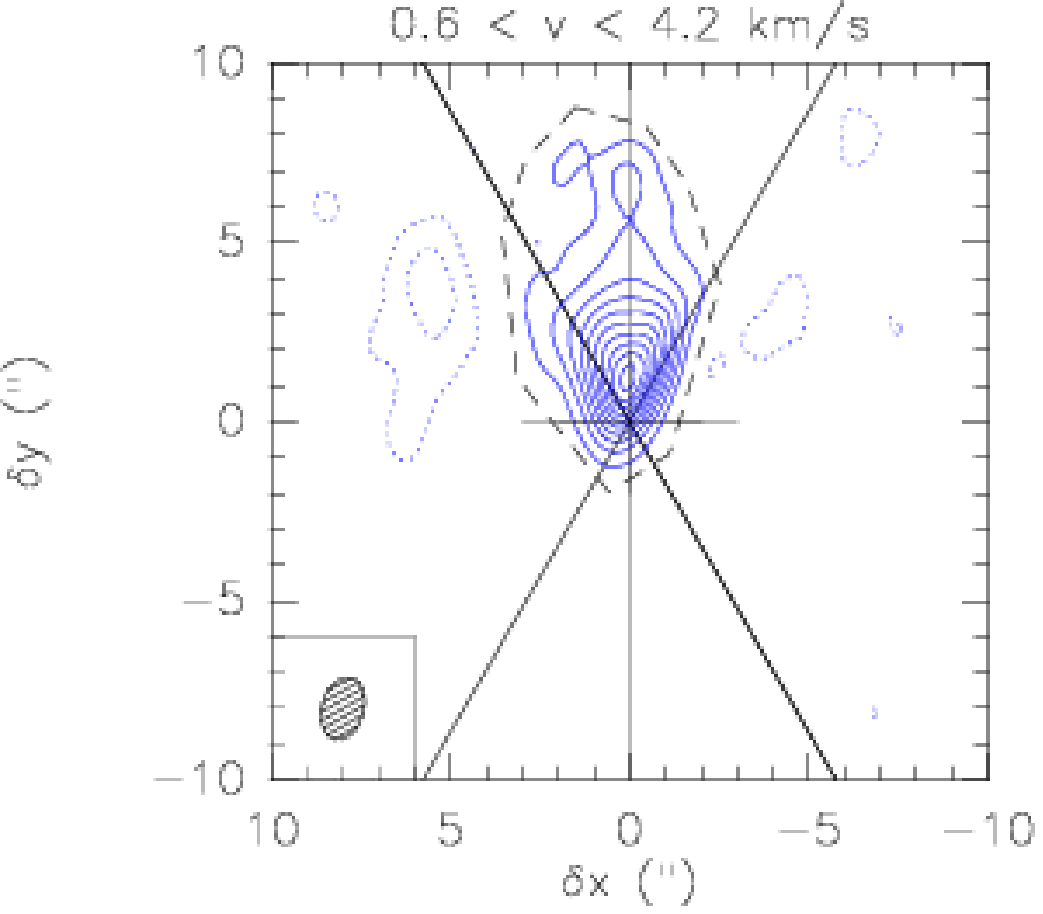}
    \includegraphics[height=0.325\hsize{},angle=270]{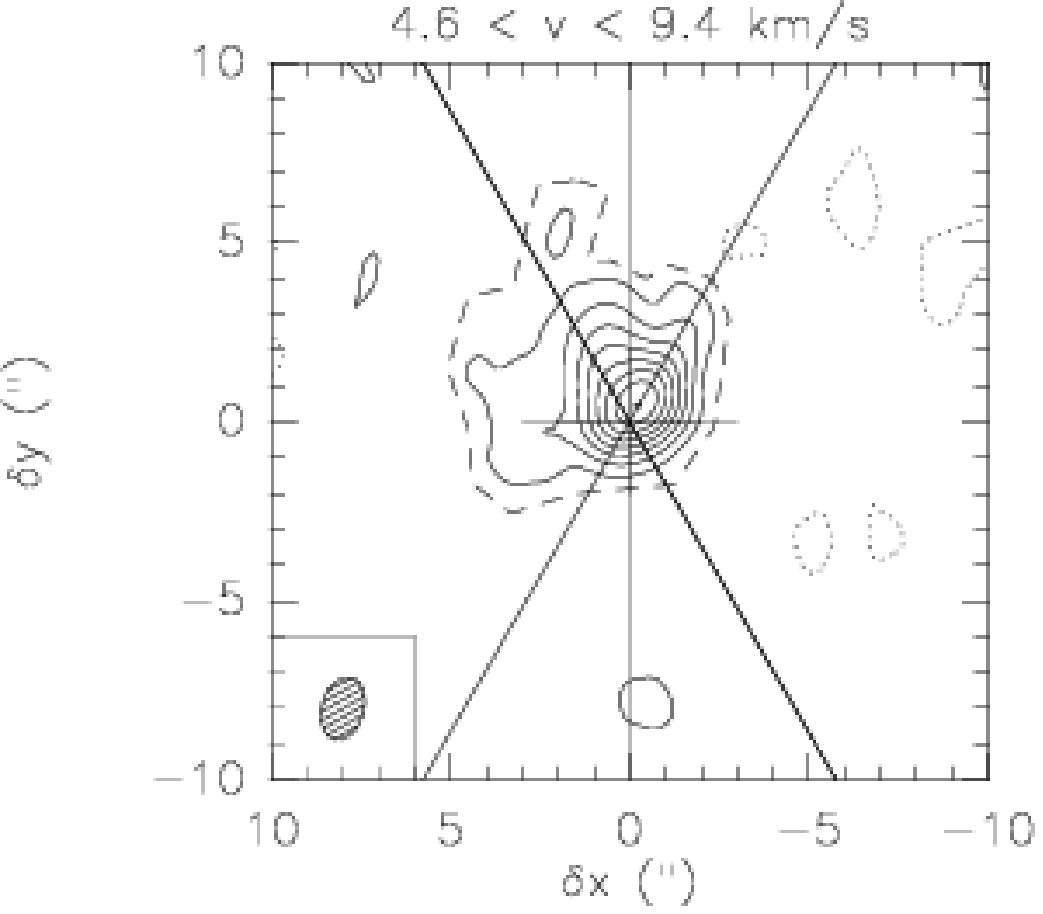}
    \includegraphics[height=0.325\hsize{},angle=270]{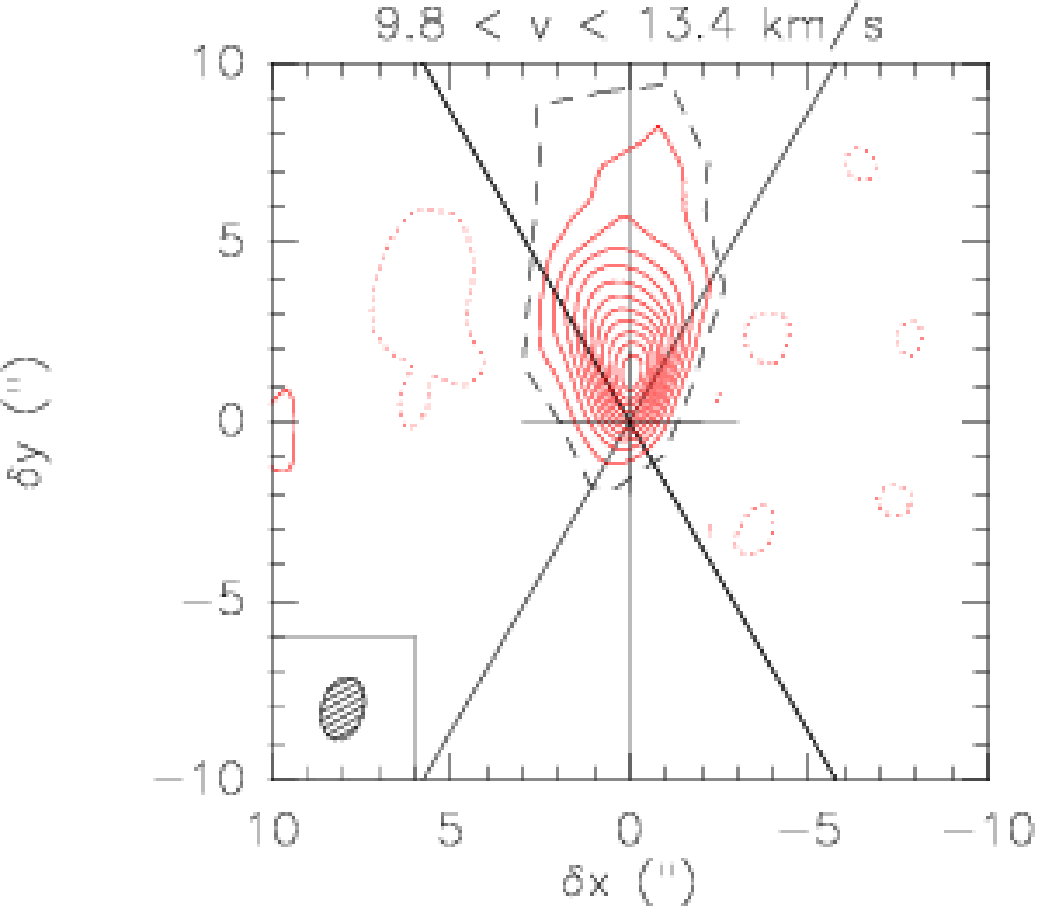}
    \caption{Integrated emissions of the \twCO{}~\Jtwo{} line in three
      contiguous velocity intervals: 1) blue--shifted velocities where only
      the outflow emits (left panel), 2) near--systemic velocities where
      confusion arises between the outflow, the disk and the cloud (medium
      panel), 3) red--shifted velocities where only the outflow emits
      (right panel). Maps have been rotated by 30\deg{} clockwise around
      the disk center to bring the optical jet direction in the vertical
      direction as this eases the extraction of position--velocity
      diagrams. This defines a new coordinate system, named $x,y$
      hereafter. In addition to the horizontal--vertical cross featuring
      the dark line and the optical jet, a second cross is sketching a cone
      of 30\deg{} half-opening angle. The contour spacing corresponds to
      3~$\sigma$ (\ie{} from left to right: 0.16, 0.18 and
      0.16~Jy/Beam\kms{}). Fluxes estimated inside polygons following the
      2~$\sigma$ contour (dashed lines) are from left to right: 6.8, 7.0
      and 8.3~Jy\kms{}.}
      \label{fig:12co21-flux}
  \end{figure*}}

\newcommand{\FigOutflowClump}{%
  \begin{figure}
    \centering %
    \includegraphics[height=\hsize,angle=270]{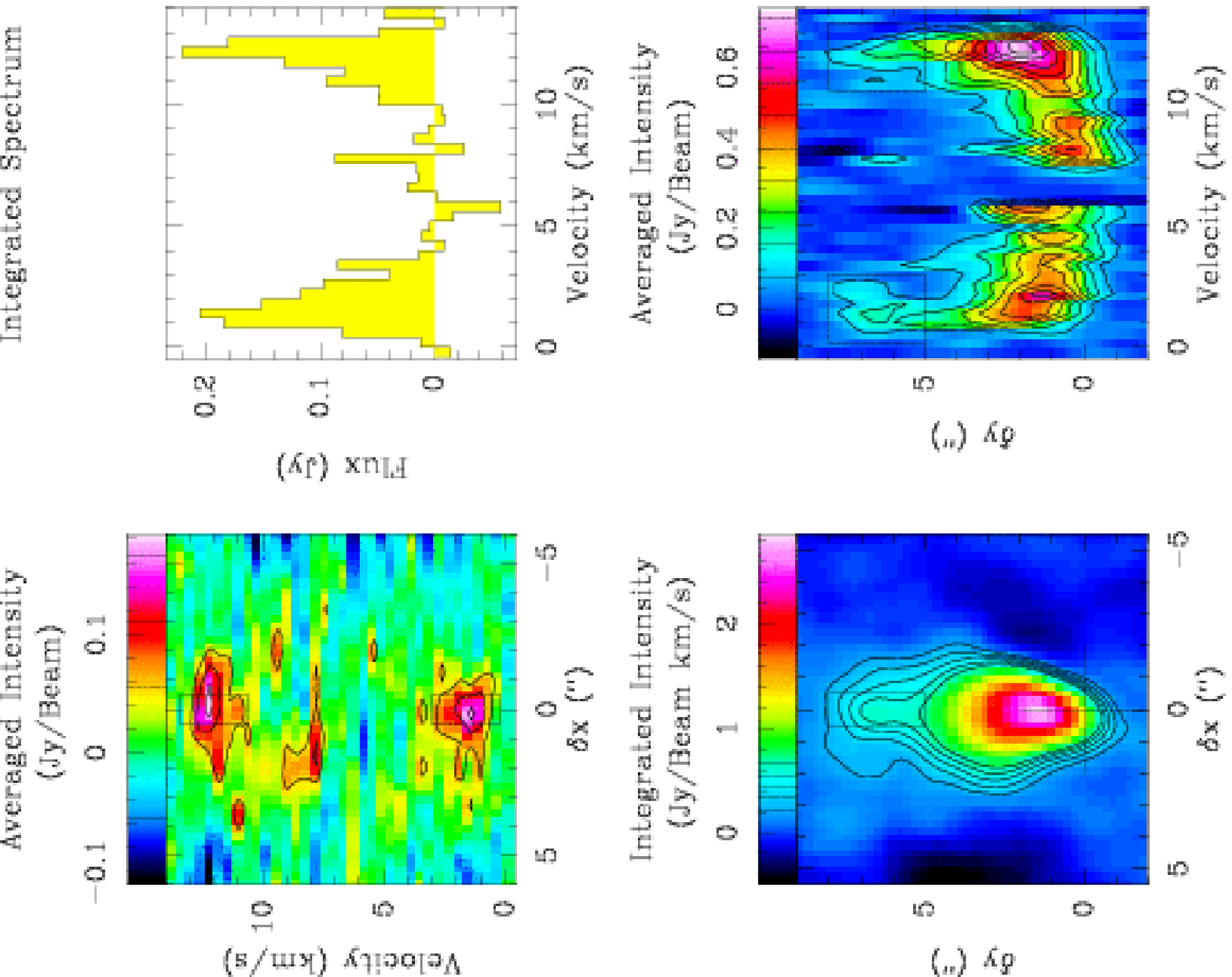}
    \caption{Spatial and kinematical properties
      of the \twCO{} \Jtwo{} ``clump'' observed along the jet axis: (a) Top
      left: $xv$ position--velocity diagram averaged on $5''< \delta y
      <8''$.  (b) bottom left: Integrated intensity in the [0,3] and
      [10.5,13.5]\kms{} velocity ranges. (c) Bottom right: $vy$
      position--velocity diagram averaged on $-0.5'' < \delta x <0.5''$.
      (d) Top right: Spectrum integrated in the $-0.5'' < \delta x <0.5''
      \times 5''< \delta y <8''$ rectangle.}
    \label{fig:clump}
  \end{figure}}

\newcommand{\FigModvsObs}{%
  \begin{figure}
    \centering %
    \includegraphics[width=\hsize{}]{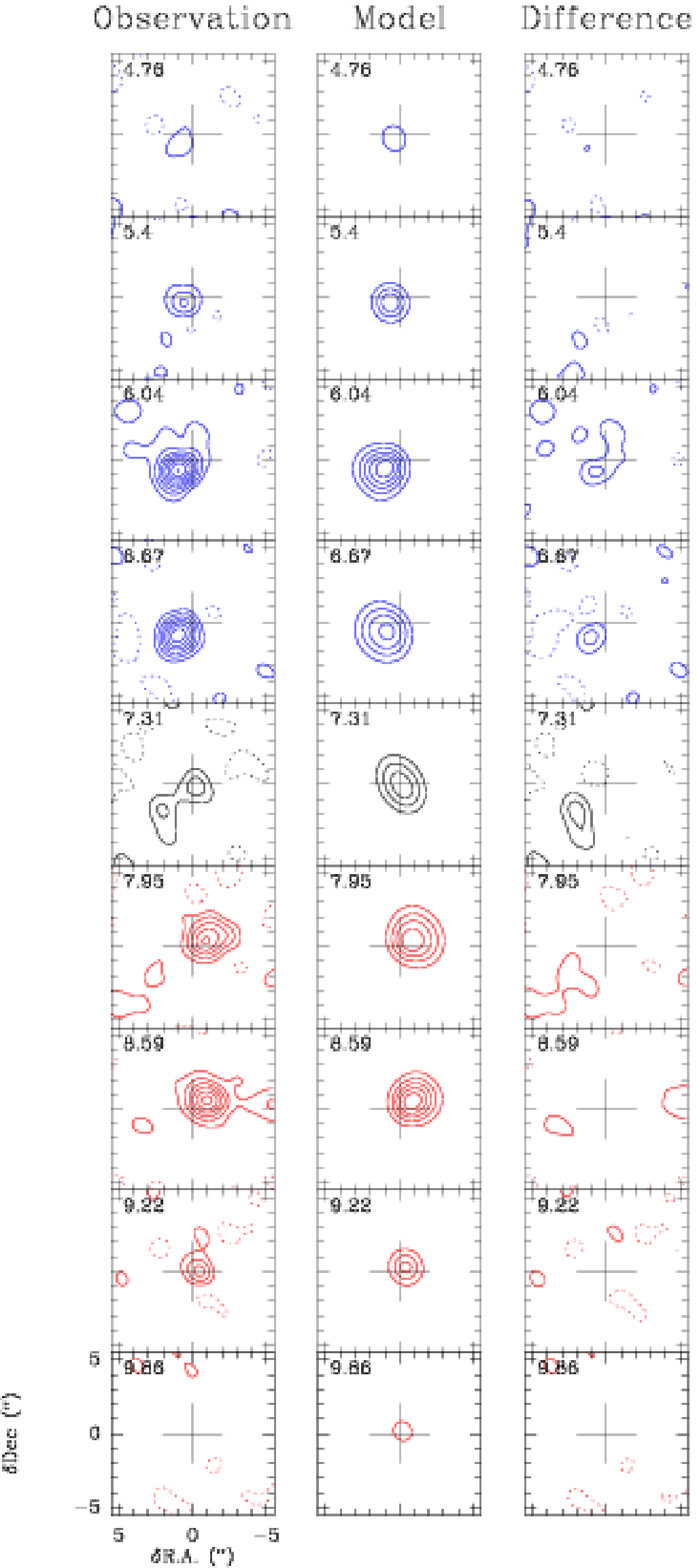}
    \caption{Channel maps of the \thCO{} \Jtwo{} emission. Observations,
      best model and difference (= observation--model) are plotted from
      left to right. The coordinate system is native (\ie{} equatorial).
      The spatial resolution is $1.8 \times 1.6"$ at PA of 26\deg{}.  All
      contours are equally spaced by a value of 45~mJy/Beam (2.5 $\sigma$
      at the spectral resolution of 0.64\kms{}).  Blue and red contours
      respectively indicate the blue and red--shifted channels while the
      black contours indicate the systemic velocity channel at 7.25\kms{}.
      Plain and dotted lines respectively show positive and negative
      contours. The cross indicates the disk center as measured in the
      1.3\mm{} continuum emission. A geometrical asymmetry
      (5$\sigma$--contours in the difference maps) is detected in the
      blue--shifted part of the disk, close to the systemic velocity.}
    \label{fig:disk:mod-vs-obs}
  \end{figure}}

\newcommand{\FigOutflowSketch}{%
  \begin{figure}
    \centering %
    \includegraphics[width=0.7\hsize{},angle=270]{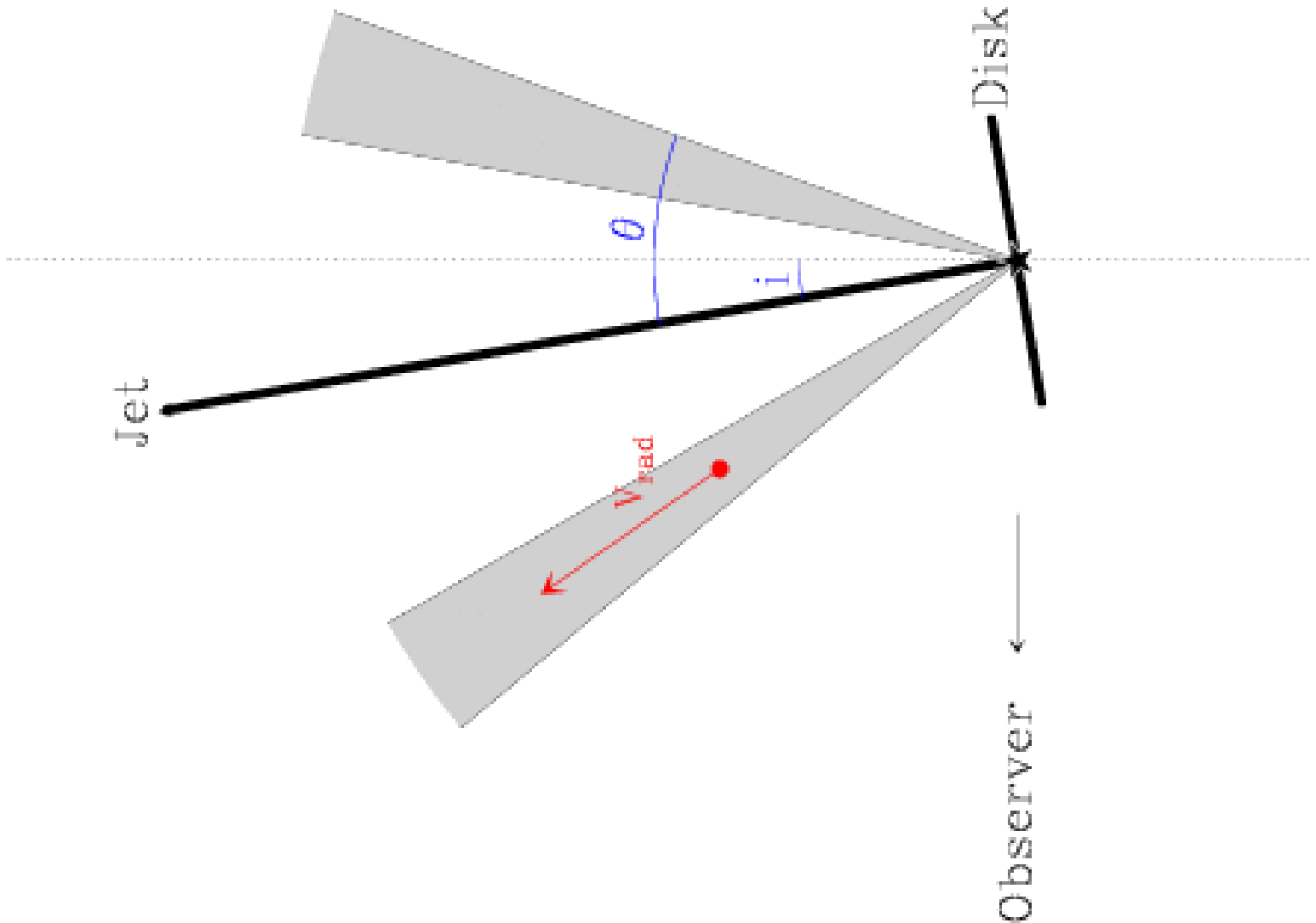}
    \caption{Sketch of our outflow model. The emitting gas is confined in a
      layer near the edge of a conical structure (grey zone). The dotted
      vertical line indicates the plane of the sky. The southern part of
      the disk is pointed toward us in agreement with the scattered light
      observations~\citep[\eg][]{burrows96}.}
    \label{fig:outflow-sketch}
  \end{figure}}

\newcommand{\FigOutflowBest}{%
  \begin{figure}
    \centering %
    \includegraphics[width=0.49\hsize{}]{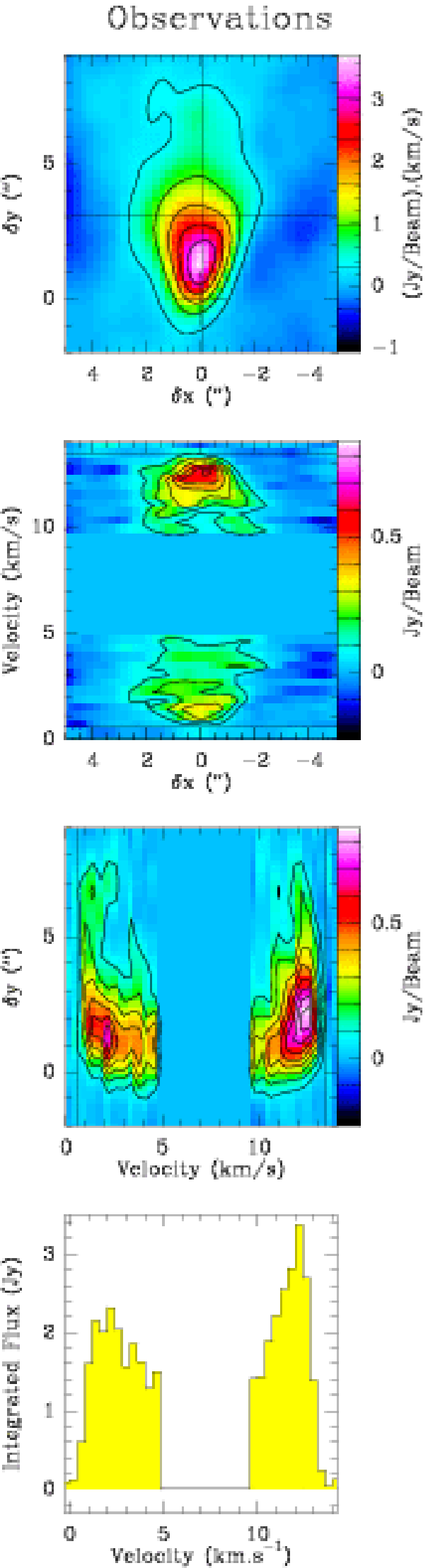}
    \hfill{}
    \includegraphics[width=0.49\hsize{}]{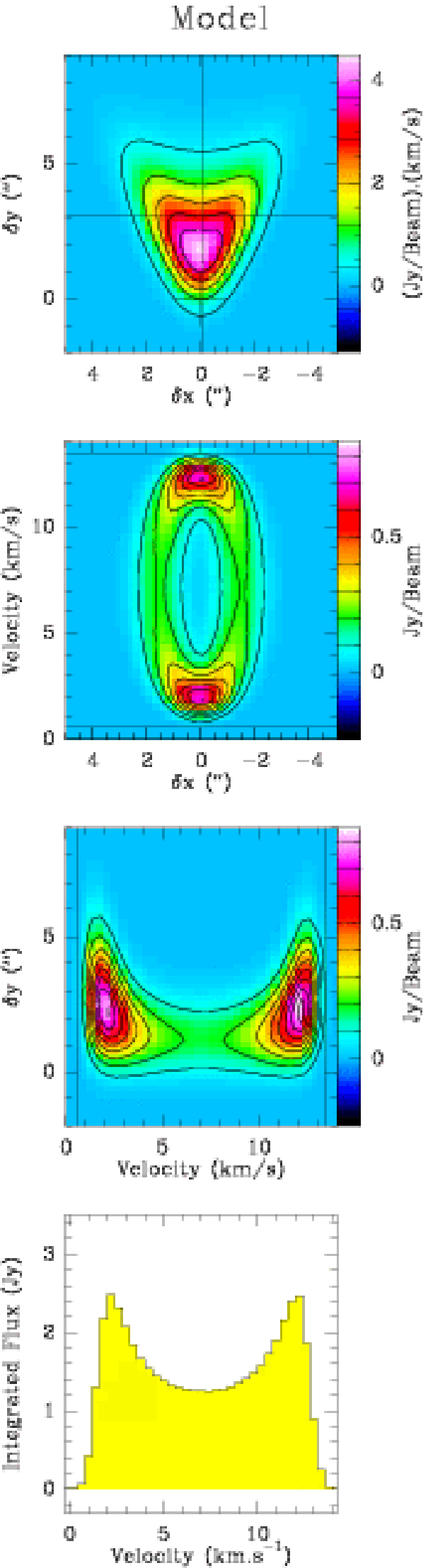}
    \caption{Comparison between \twCO{}~\Jtwo{} observations (left column)
      and our best model (right column) of the outflow. The velocity
      channels where confusion from the disk or the parent cloud exist have
      been flagged in the \twCO{}~\Jtwo{} data cube. The integrated
      emission map, the position--velocity diagrams perpendicular and
      parallel to the jet axis and the spectrum integrated over the map are
      shown from top to bottom. The plots are shown in the rotated ($\delta
      x, \delta y$) coordinate system.  The horizontal and vertical lines
      on the integrated emission map show the position of the cuts used to
      form the position--velocity diagrams (respectively $\delta x =
      -0.1''$ and $\delta y = 3.1''$). The two horizontal or vertical lines
      in the position--velocity diagrams indicate the velocity range over
      which the emission have been integrated to form the top map ($0.6 < V
      < 13.4\kms$).}
  \label{fig:outflow:mod-vs-obs}
  \end{figure}}

\newcommand{\FigOutflowXY}{%
  \begin{figure*}
    \centering %
    \includegraphics[height=0.475\hsize{},angle=270]{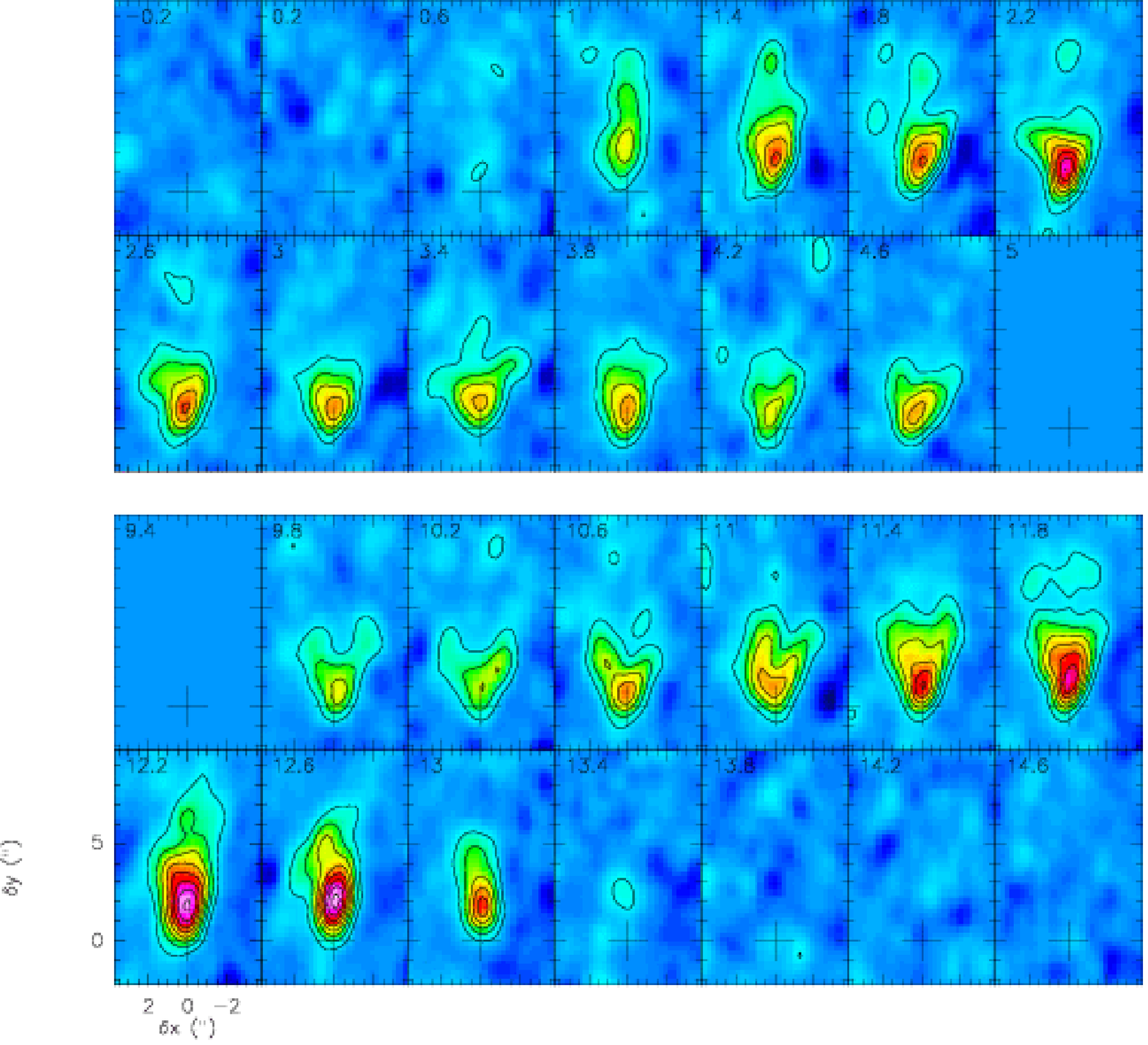}
    \includegraphics[height=0.475\hsize{},angle=270]{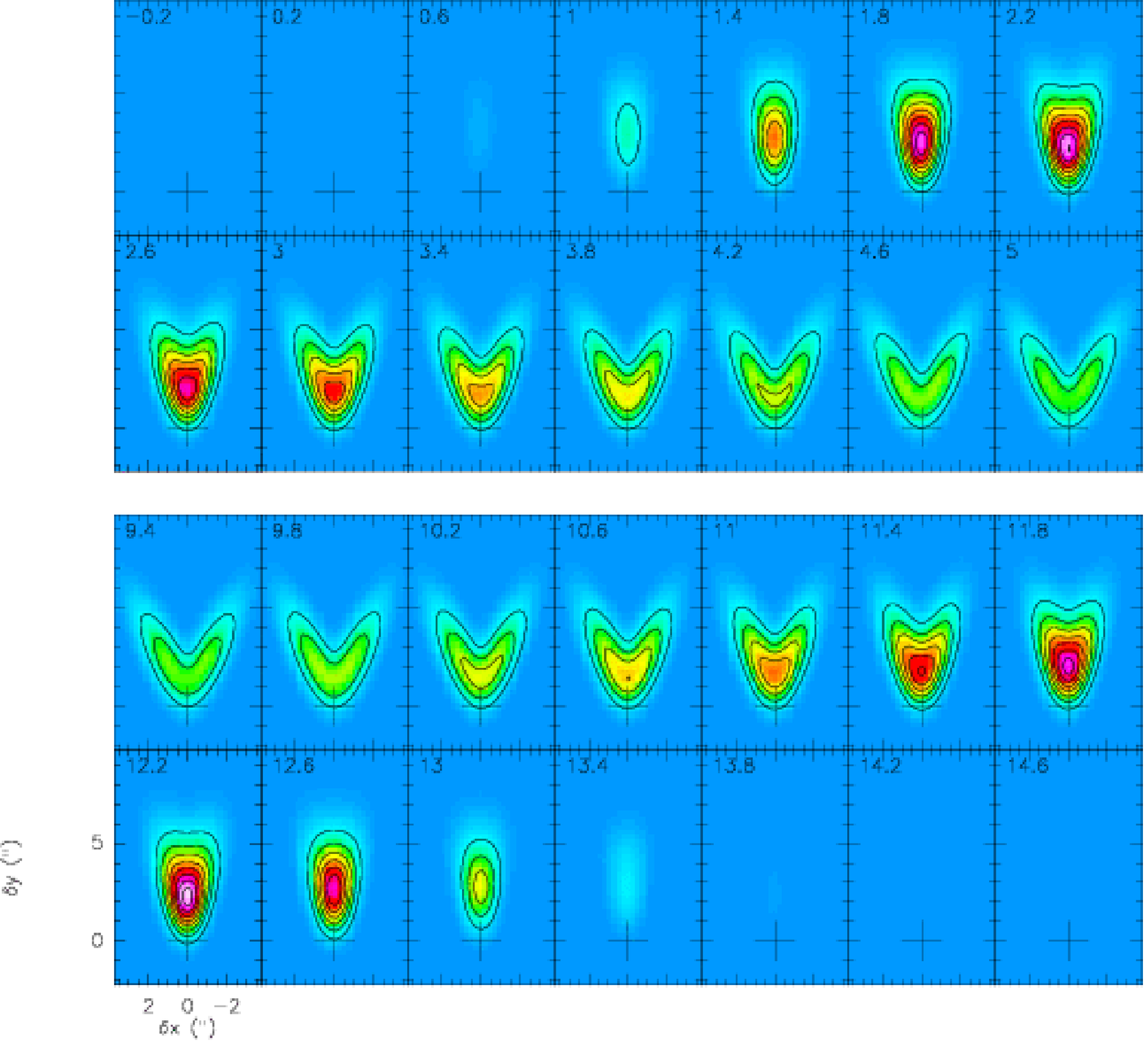}
    \caption{Comparison of the \emph{channel maps} of the \twCO{}~\Jtwo{} 
      observations (left) and our best model (right) of the outflow. The
      cross indicates the disk center as measured in the 1.3\mm{} continuum
      emission. Velocity of the channels are written in\kms{} on the top
      left corner of each channel. Both channel maps share the same color
      scale and contour spacing (0.1~Jy/beam).}
  \label{fig:outflow:xy}
  \end{figure*}}

\newcommand{\FigOutflowXV}{%
  \begin{figure*}
    \centering %
    \includegraphics[height=0.475\hsize{},angle=270]{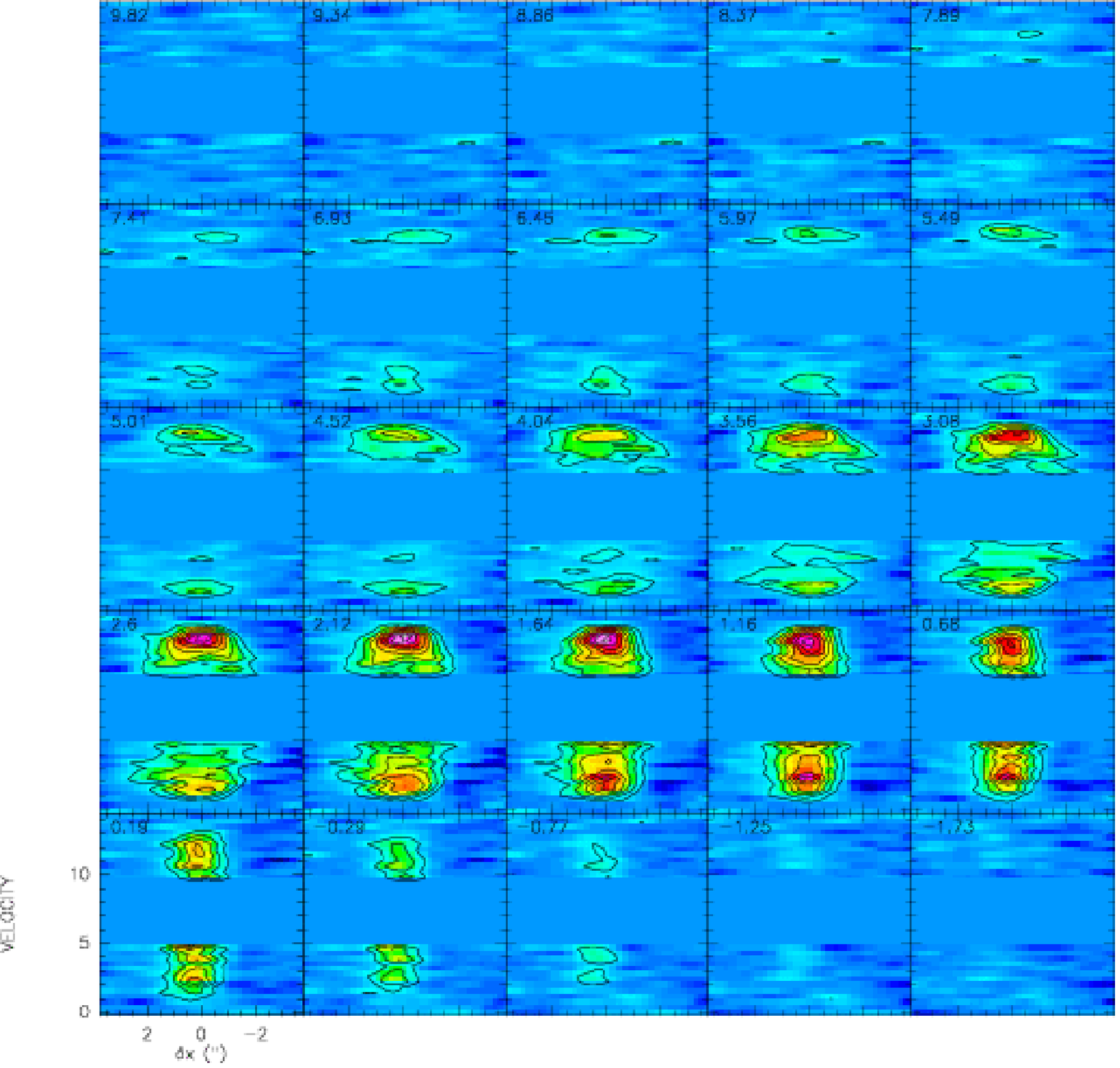}
    \includegraphics[height=0.475\hsize{},angle=270]{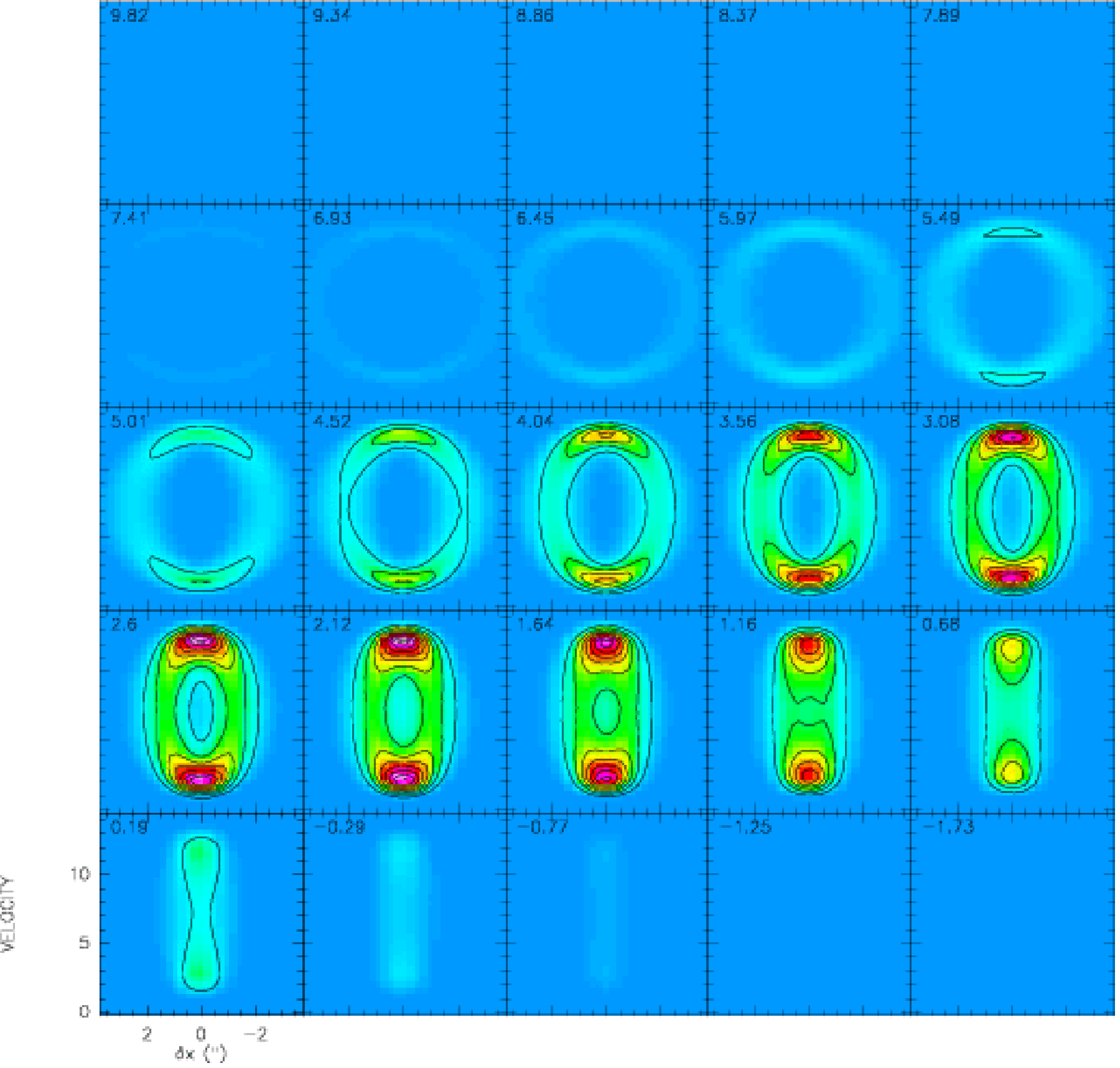}
    \caption{Comparison of the \emph{position--velocity diagrams 
        perpendicular to} the jet axis of the \twCO{}~\Jtwo{} observations
      (left) and our best model (right) of the outflow.  The coordinate
      along the jet axis ($\delta y$) are written in arcsecond on the top
      left corner of each panel. All the diagrams share the same color
      scale and contour spacing (0.1~Jy/beam).}
  \label{fig:outflow:xv}
  \end{figure*}}

\newcommand{\FigOutflowVY}{%
  \begin{figure*}
    \centering %
    \includegraphics[height=0.475\hsize{},angle=270]{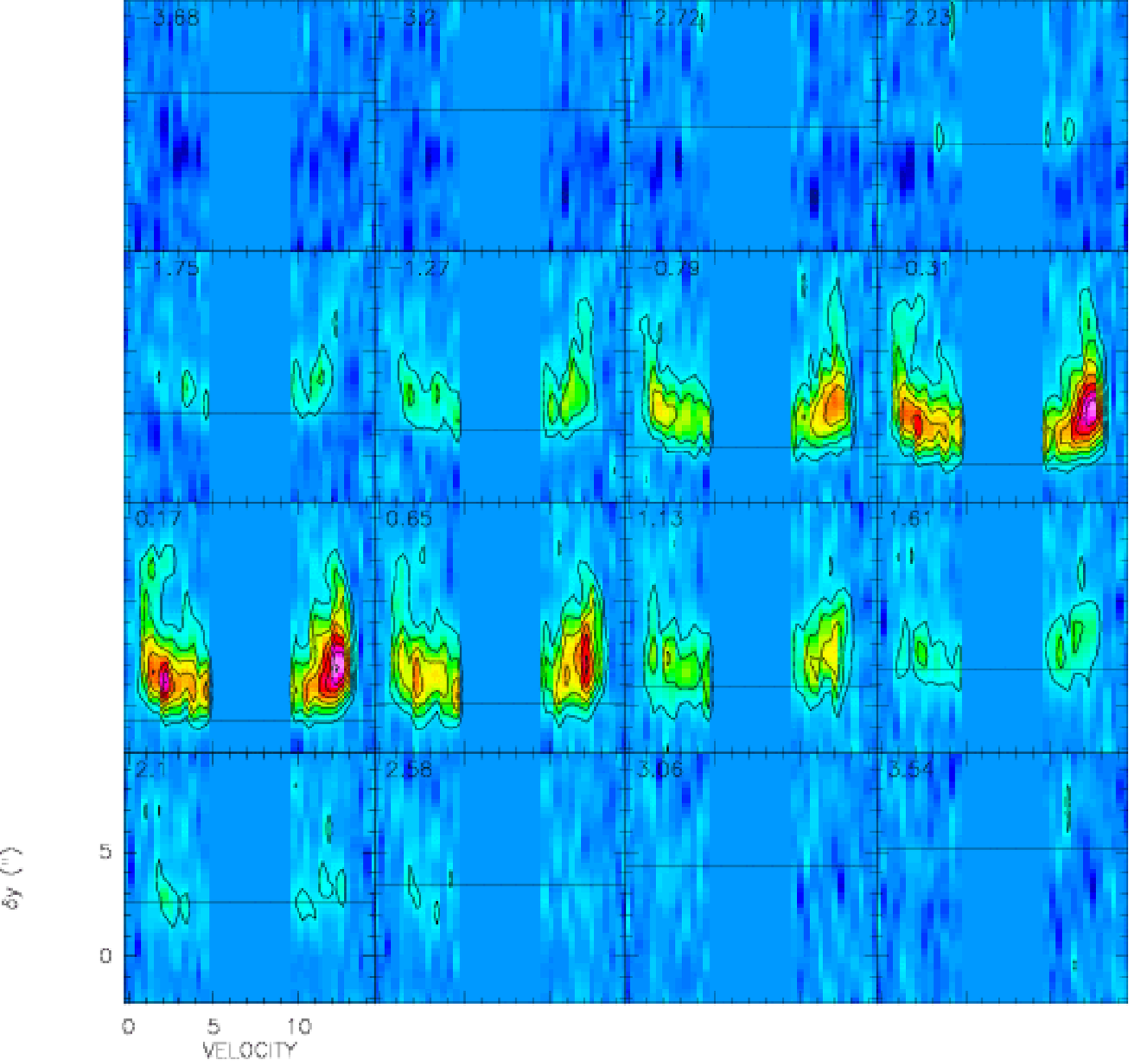}
    \includegraphics[height=0.475\hsize{},angle=270]{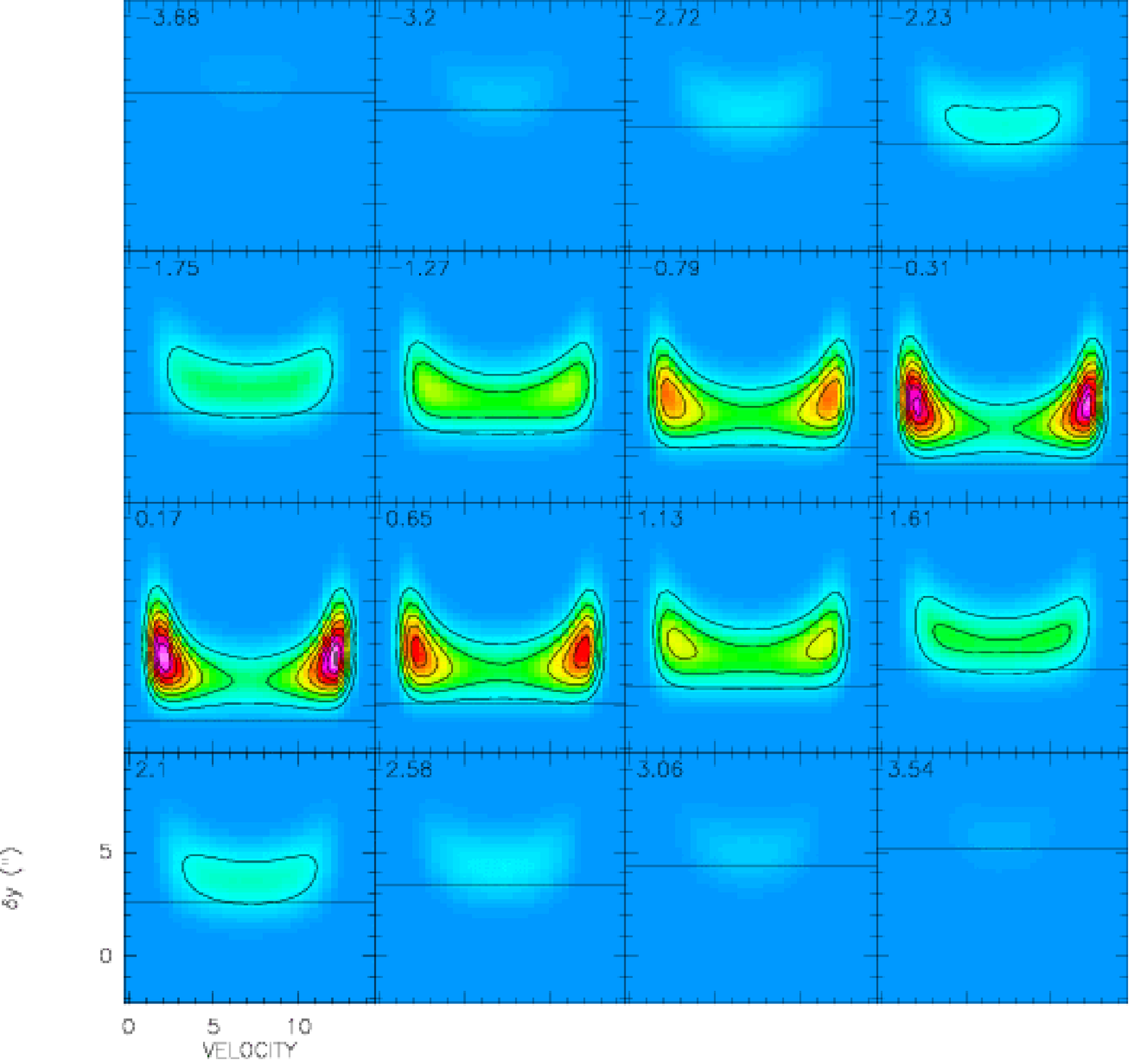}
    \caption{Comparison of the \emph{position--velocity diagrams along} 
      the jet axis of the \twCO{}~\Jtwo{} observations (left) and our best
      model (right) of the outflow.  The coordinate perpendicular to the
      jet axis ($\delta x$) are written in arcsecond on the top left corner
      of each panel. The horizontal line indicates the position of the
      outside edge of a cone of 30\deg{} half opening--angle. All the
      diagrams share the same color scale and contour spacing
      (0.1~Jy/beam).}
  \label{fig:outflow:vy}
  \end{figure*}}

\newcommand{\FigOutflowParaOne}{
  \begin{figure*}
    \centering %
    \includegraphics[width=0.24\hsize{}]{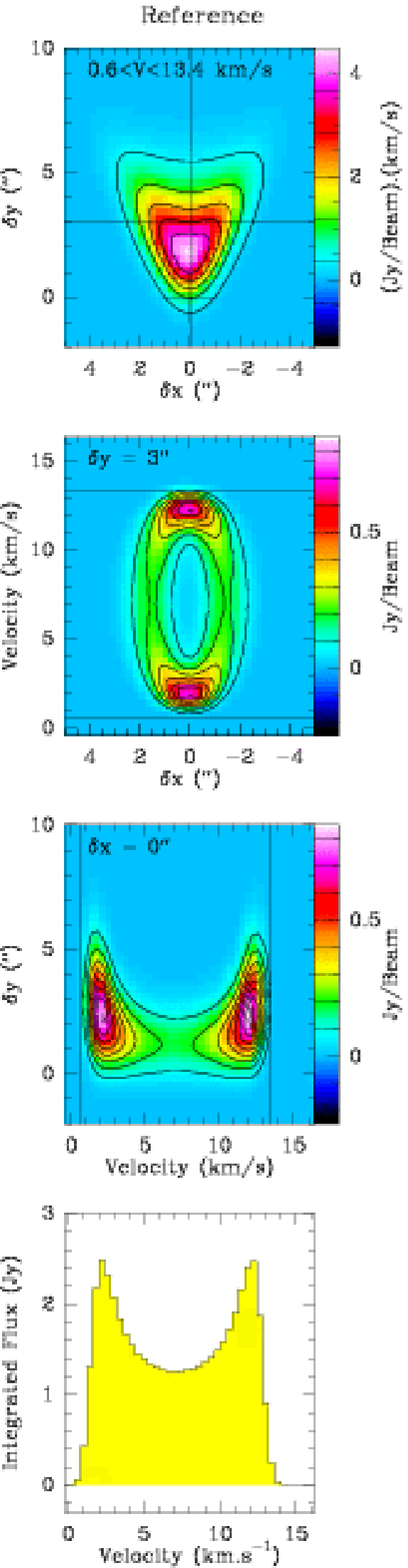}
    \hspace{0.005\hsize{}}
    \includegraphics[width=0.24\hsize{}]{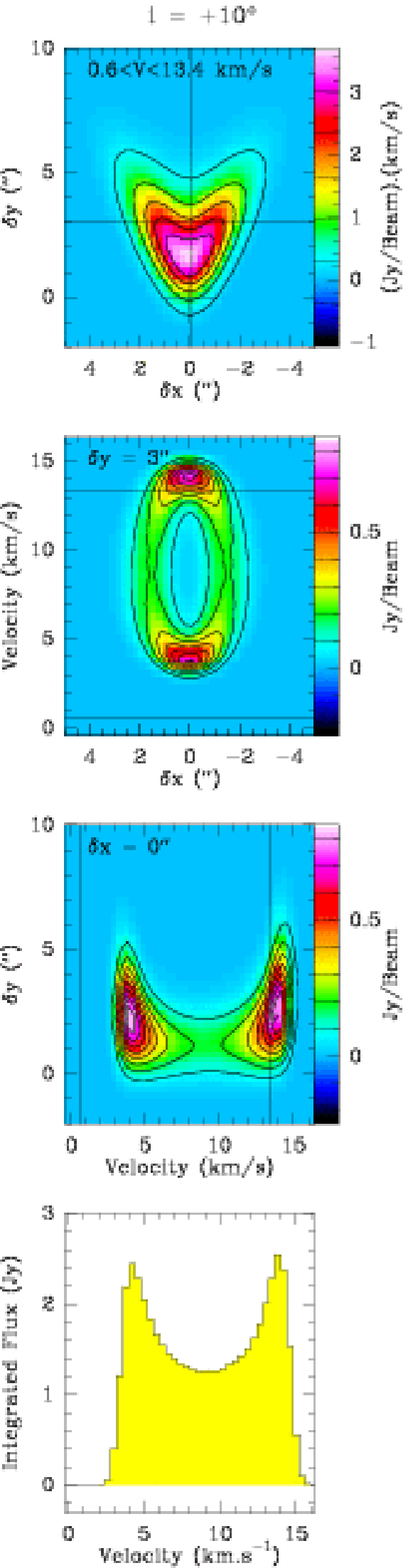}
    \hspace{0.005\hsize{}}
    \includegraphics[width=0.24\hsize{}]{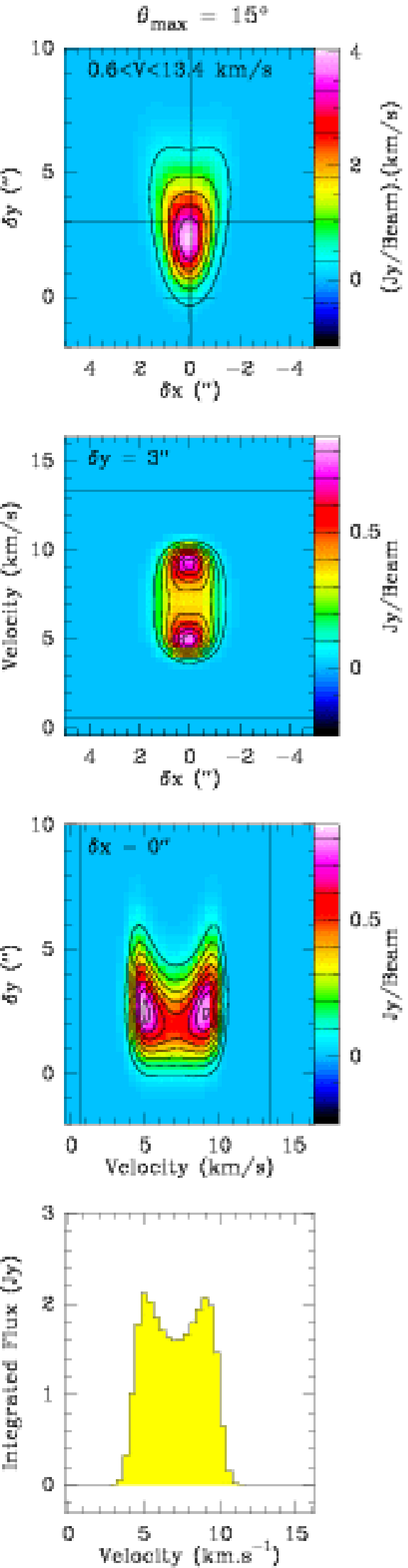}
    \hspace{0.005\hsize{}}
    \includegraphics[width=0.24\hsize{}]{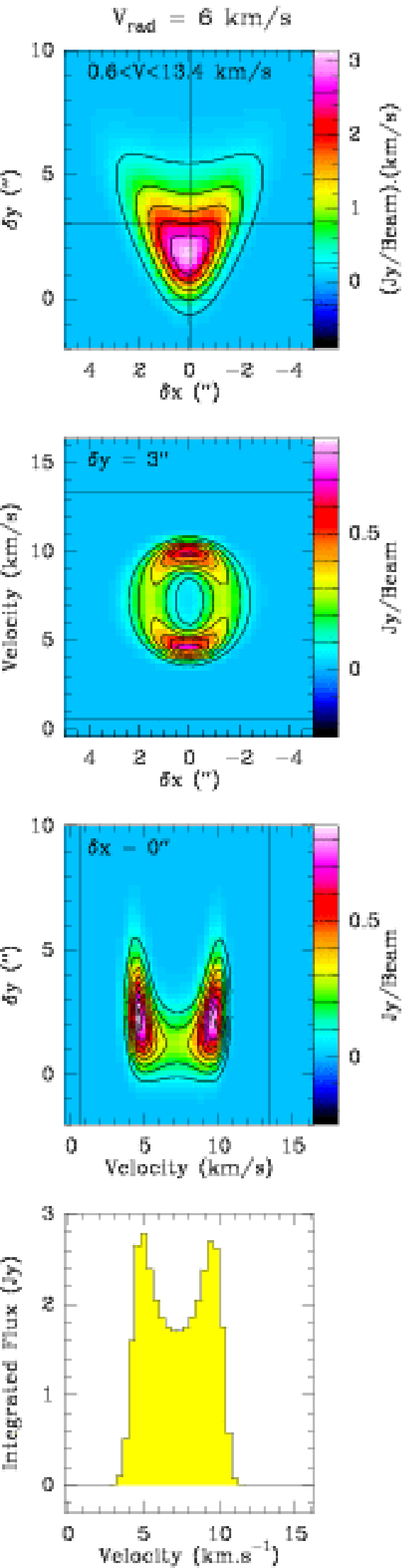}
    \caption{Influence of the different parameters of the outflow model. 
      The left column shows our reference model whose parameters are given
      in Table~\ref{tab:outflow}. The each column shows a model for which
      one and only one parameter (noted on the column top) has been varied
      compared to our reference model. The integrated emission map, the
      position--velocity diagrams perpendicular and parallel to the jet
      axis and the spectrum integrated over the map are shown from top to
      bottom.}
  \label{fig:outflow:para:1}
  \end{figure*}}

\newcommand{\FigOutflowParaTwo}{
  \begin{figure*}
    \centering %
    \includegraphics[width=0.24\hsize{}]{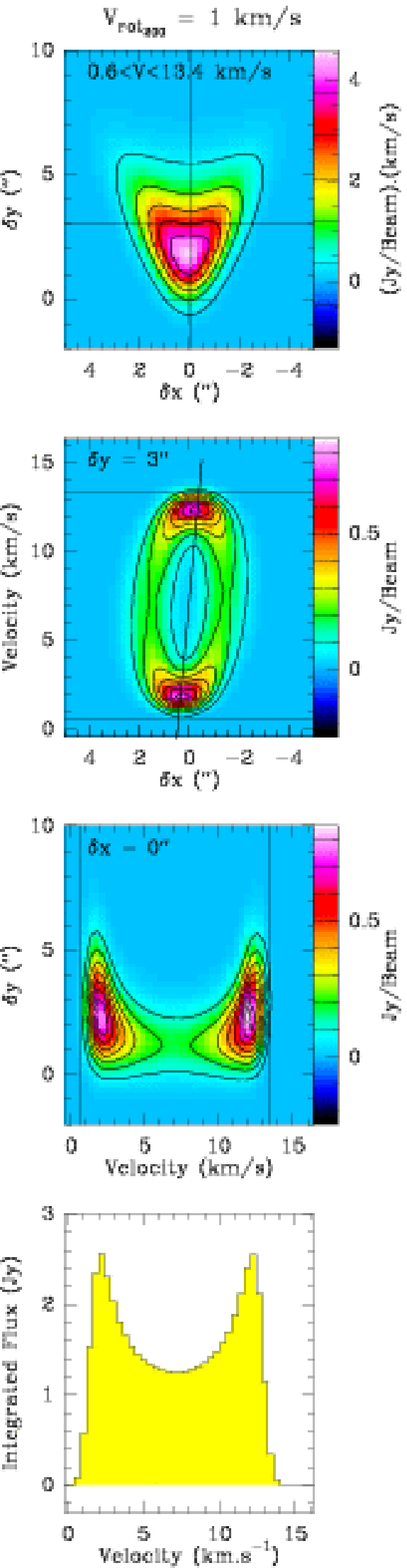}
    \hspace{0.005\hsize{}}
    \includegraphics[width=0.24\hsize{}]{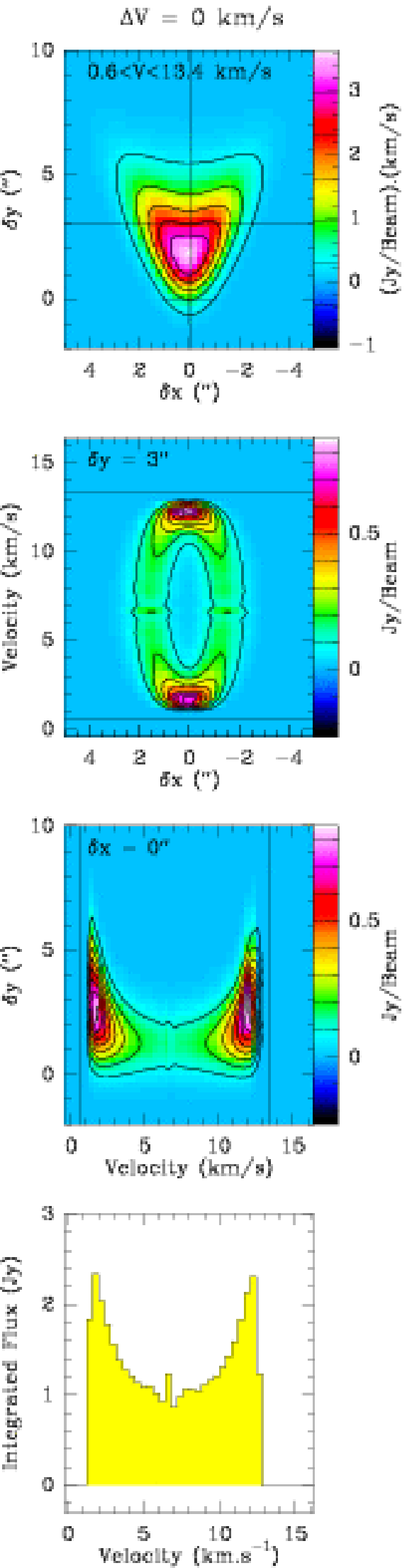}
    \hspace{0.005\hsize{}}
    \includegraphics[width=0.24\hsize{}]{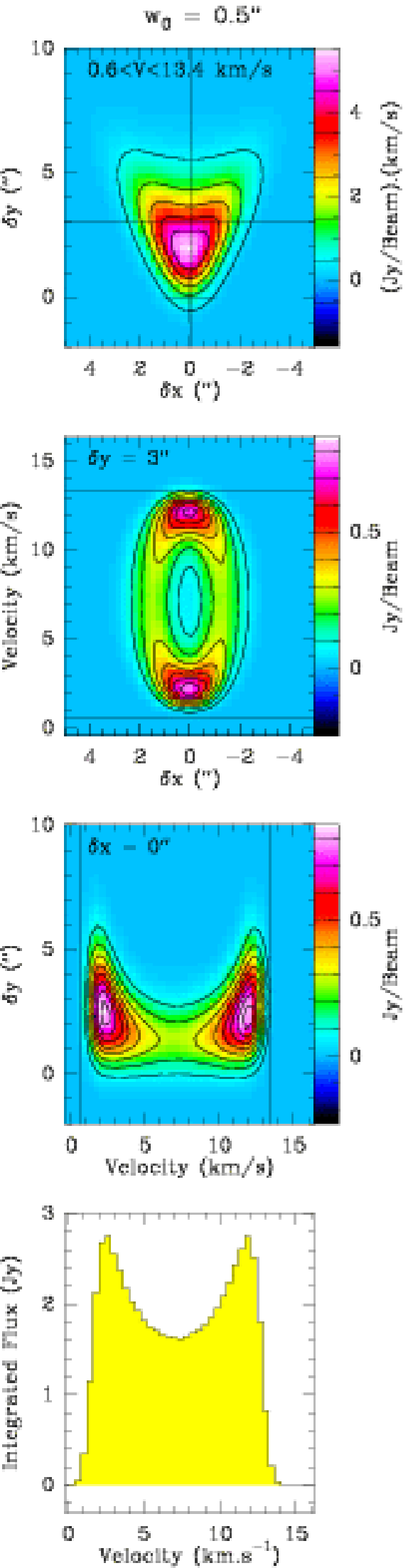}
    \hspace{0.005\hsize{}}
    \includegraphics[width=0.24\hsize{}]{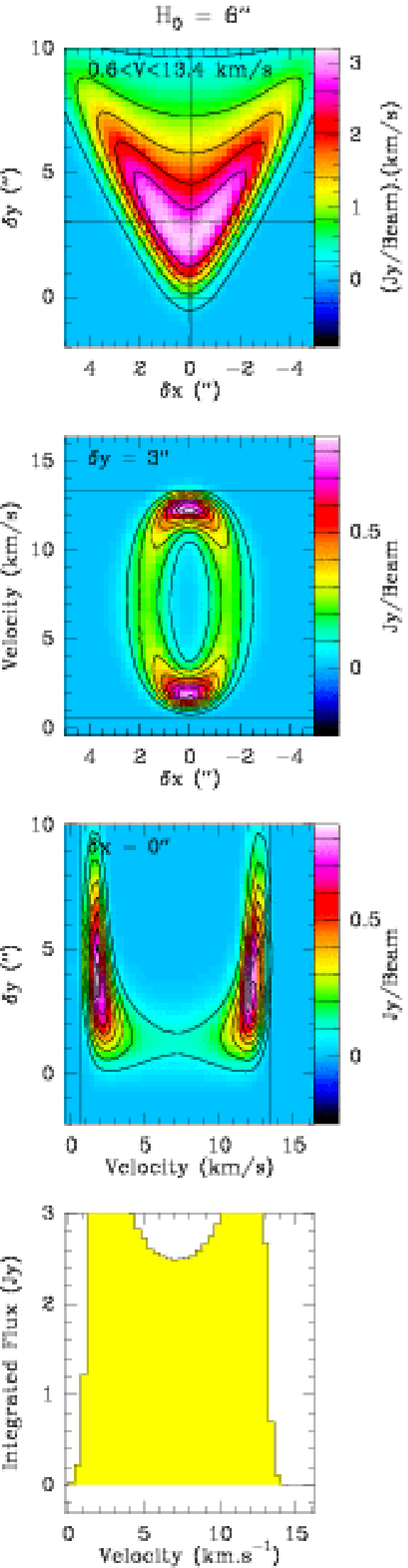}
    \caption{Continuation of Fig.~\ref{fig:outflow:para:1} with different
      varying parameters.}
  \label{fig:outflow:para:2}
  \end{figure*}}

\begin{document}

\title{Plateau de Bure Interferometer Observations \\
  of the Disk and Outflow of \HHth{}\thanks{PdBI is operated by IRAM, which
    is supported by INSU/CNRS (France), MPG (Germany), and IGN (Spain).}}

\author{J.~Pety\inst{1,2}%
  \and F.~Gueth\inst{1}%
  \and S.~Guilloteau \inst{3}%
  \and A.~Dutrey \inst{3}}

\institute{%
  IRAM, 300 rue de la Piscine, 38400 Saint Martin d'H\`eres, France.\\
  \email{pety@iram.fr, gueth@iram.fr}%
  \and{} LERMA, UMR 8112, CNRS, Observatoire de Paris, 61 av. de
  l'Observatoire,
  75014, Paris, France.%
  \and{} L3AB, UMR 5804, Observatoire de Bordeaux, 2 rue de l'Observatoire, 
  33270 Floirac, France.\\
  \email{Stephane.Guilloteau@obs.u-bordeaux1.fr,
    Anne.Dutrey@obs.u-bordeaux1.fr}}

\offprints{J. Pety, \email{pety@iram.fr}}

\date{Received 13 June 2006, Accepted 26 July 2006} %
\authorrunning{Pety et al.} %

\abstract
{\HHth{} is a well-known Pre-Main-Sequence star in Taurus.  HST
  observations have revealed a flared, edge-on disk driving a
  highly-collimated optical jet, making this object a case study for the
  disk-jet-outflow paradigm.}
{We searched for a molecular outflow, and attempted to better constrain the
  star and disk parameters.}
{We obtained high angular resolution ($\about{}1''$) observations of the
  dust continuum at 2.7 and 1.3~mm, and of the \twCO{}~\Jtwo{},
  \thCO{}~\Jtwo{} \&~\Jone{}, \CeiO{}~\Jone{}~emissions around \HHth{}. A
  standard disk model is used to fit the \thCO{} \Jtwo{} $uv$-plane
  visibilities and derive the disk properties, and the stellar mass. An
  \emph{ad hoc} outflow model is used to reproduce the main properties of
  the \twCO{}~\Jtwo{} emission.}
{The rotation vector of the disk points toward the North-Eastern jet. The
  disk rotation is Keplerian: Using a distance of 140\pc{}, we deduce a
  mass of 0.45\Msun{} for the central star. The disk outer radius is
  420\au{}. A highly asymmetric outflow originates from the inner parts of
  the disk. Only its North-Eastern lobe was detected: it presents to first
  order a conical morphology with a $30\degr{}$ half opening angle and a
  constant (12\kms{}) radial velocity field. Outflow rotation was searched
  for but not found. The upper limit of the outflow rotation velocity is
  1\kms{} at 200\au{} of the jet axis.}
{\HHth{} is a low mass TTauri of spectral type around M1 and age 1 to
  4~Myrs, surrounded by a medium size Keplerian disk, of mass around $4\,
  10^{-3}\Msun$. It beautifully illustrates the jet-disk-outflow
  interaction, being so far the only star to display a jet and outflow
  connected to a well defined Keplerian disk, but reveals a surprisingly
  asymmetric (one-sided) outflow despite a relatively symmetric jet.
  Furthermore, these observations do not enable to assign the origin of the
  molecular outflow to entrainment by the optical jet or to a disk wind. In
  the latter hypothesis, the lack of rotation would imply an origin in the
  inner 15\au{} of the disk.}


\maketitle %

\section{Introduction}

\TableObs{} %

During the last decade, the properties of the circumstellar environment of
pre-main-sequence (PMS) stars have been extensively studied. Millimeter
arrays, such as the IRAM Plateau de Bure interferometer, routinely provide
CO line and continuum images of low-mass PMS stars which show that many of
them are surrounded by large ($R_\emr{out}\sim 200-800\au$) disks in
Keplerian rotation
\citep[\eg{}][]{koerner93,dutrey94,mannings97,guilloteau98,simon00}. For
several sources, the angular resolution and sensitivity of the observations
enable to fit disk models to the data, hence constraining physical
parameters of the disk, and even the mass of the central
source~\citep[][and references therein]{dutrey06}. While mm continuum
observations are sensitive to the dust thermal emission of the disk,
near-infrared (NIR) or optical observations can be used to trace the
stellar light scattered by particles at the disk surface
\citep[\eg{}][]{roddier96}. If the system is edge-on, the disk itself can
be seen as a dark lane on the equatorial plane (\eg{}
\HHth{}:~\citet{burrows96}, GM Auriga:~\citet{stapelfeldt97}).

Most of pre-main-sequence stars also show an ejection of matter.  In
several sources, optical linear jets are observed through the emission of
forbidden atomic lines. Interestingly, \citet{bacciotti03} and
\citet{coffey04} have detected possible rotation signatures in at least
four jets (DG\,Tau, TH\,28, RW\,Aur, LkH$\alpha$\,321): They have observed
systematic velocity differences between opposite sides of the jet axis.
Indeed, rotation of the jet is predicted by magnetocentrifugal models of
accretion-ejection~\citep{konigl00,shu00}, through the coupling of the
outflowing gas to the helicoidal structure of the magnetic field. In
younger, more embedded sources, large molecular outflows are observed,
mainly through the emission of the rotational lines of CO and its
isotopologues~\citep[][and references therein]{arce06}. Those molecular
outflows are usually identified with ambient molecular gas that has been
entrained and thus put into motion by the underlying protostellar jet. In
any case, it is particularly interesting to check whether a rotation
component is present in molecular outflows.

In this context, one of the most interesting objects is the young PMS star
\HHth{}, located in the Taurus molecular cloud at a distance of $\sim
140\pc$.  \citet{burrows96} obtained spectacular images with the Hubble
Space Telescope that revealed a flared edge-on disk of diameter $\sim
450\au$, separated in two parts by a dust lane, and a highly-collimated jet
emanating out of the central region. Since then, the \HHth{} disk and jet
have been studied in many details \citep[\eg{}][]{bacciotti99,
  stapelfeldt99a, wood00, cotera01, wood02, watson04}.
\citet{stapelfeldt99a} reported variability and asymmetry in the emission
from the circumstellar disk by comparing HST optical observations performed
between 1994 and 1998; the photometric variability of the star and its
possible effects on the disk was investigated by~\citet{wood00}. This
source is a unique candidate to probe the complex interaction between the
disk, the jet and the outflow.

However, \HHth{} has been poorly investigated at mm wavelengths so far.
This is probably due to the confusion arising from the \HHth{} parent
molecular cloud, which increases the complexity of any detailed
observations using the emission of CO isotopologues. The \twCO{} emission
of the cloud is expected to be optically thick, hence hiding significantly
the disk emission. BIMA observations of the \thCO{}~\Jone{} emission in the
environment of HL~Tau~\citep[Fig.~2 of][]{welch00} shows that \HHth{} is
located in the south edge of a shell driven by XZ~Tau. Still, the
remarkable properties of \HHth{} makes it a textbook case for the study of
the disk/jet system in young PMS stars, and we have therefore decided to
use the IRAM Plateau de Bure interferometer to perform high angular
resolution ($\about{}1''$) observations of the emission of the
circumstellar material surrounding \HHth{} in several CO isotopologues
lines. In this paper we present the results of this study and the analysis
of the thermal dust and molecular line emissions of the disk and the
outflow.

\section{Observations and data reduction}

\subsection{\twCO{}, \HCOp{} and continuum at 3.4 and 1.3~mm}

First PdBI observations dedicated to this project were carried out with 5
antennas in BCD configurations (baseline lengths from 24 to 280~m) during
the winter 1997/1998. The observations were performed simultaneously at
89.2~GHz (\HCOp{}~\Jone{}) and 230.5~GHz (\twCO{}~\Jtwo{}).  One correlator
band of 10 MHz was centered on the \HCOp{}~\Jone{}~line. Another band of 20
MHz was centered on the \twCO{}~\Jtwo{}~line. Finally, two bands of 160 MHz
were used for the 1.3~mm and 3.4~mm continuum, respectively.

Those observations were part of the survey of protoplanetary disks which is
described by~\citet{simon00}. The interferometer was operating in track
sharing mode, \ie{} each 8h--track was shared between several close--by
sources. The \emph{total on--source} observing time dedicated to \HHth{}
was about 6~hrs. The rms phase noises were between 8 and 25\deg{} at 3.4~mm
and between 15 and 50\deg{} at 1.3~mm.  This introduced position errors of
$<0.1''$. The seeing, estimated from observations of the calibrators, was
$\sim 0.3''$.  Typical resolutions are $3.6''$ at 3.4~mm and $1.4''$ at
1.3~mm.

\subsection{\thCO{}, \CeiO{} and continuum at 2.7 and 1.35~mm}

As a follow-up, we carried out observations of \thCO{} and \CeiO{} at PdBI
with 5 antennas in CD configuration (baseline lengths from 24 to 176~m) in
December 2000, February 2001 and March 2002. The \CeiO{}~\Jone{},
\thCO{}~\Jone{} and \thCO{}~\Jtwo{} lines were observed simultaneously
using the 3~mm receiver (tuned at 109.9~GHz) and the 1~mm receivers (tuned
at 220.4~GHz). Three 20~MHz correlator bands were centered on the
\CeiO{}~\Jone{}, \thCO{}~\Jone{} and \thCO{}~\Jtwo{} lines.  Two bands of
320 MHz were used for the 1.35~mm and 2.7~mm continuum, respectively.

The rms phase noises were between 10 and 25\deg{} at 2.7~mm and between 20
and 50\deg{} at 1.35~mm. The \emph{total on--source} observing time was
about 14~hrs. Typical resolutions are $3''$ at 2.7~mm and $1.7''$ at
1.35~mm.

\subsection{Data reduction}

All data were reduced using the \textsc{gildas}\footnote{See
  \texttt{http://www.iram.fr/IRAMFR/GILDAS} for more information about the
  \textsc{gildas} softwares.} softwares supported at IRAM~\citep[]{pety05}.
Standard calibration methods using close calibrators were applied to all
the PdBI data. Images were produced using natural weighting of the
visibilities for the line maps and robust weighting for the continuum maps.

The 1.30~mm and 1.35~mm continuum observations were merged (after flux
correction using a spectral index of 2.22, see Section~\ref{sec:data:cont})
to produce a map of higher signal--to--noise ratio in
Fig.~\ref{fig:pdbi-on-hst}. The channel maps presented in
Figs.~\ref{fig:13co-c18o} and~\ref{fig:12co21} include the continuum
emission because it is \emph{in this case} weak compared to the line
emission. It has nonetheless been subtracted before fitting the disk (cf.\ 
section~\ref{sec:disk:analysis}).

\section{Results}

\FigObsSum{} %
\FigIsotCO{} %
\FigMainCO{} %
\TableContinuum{} %

All observed lines were detected although with very different strengths.
\twCO{}~\Jtwo{}, \thCO{}~\Jone{} and \thCO{}~\Jtwo{} emissions are strong
while only weak \CeiO{}~\Jone{} and \HCOp{}~\Jone{} emissions are observed.
Continuum emission at 3.4~mm, 2.7~mm, 1.35~mm and 1.30~mm is detected, the
source being resolved only at 1.30~mm. Fig.~\ref{fig:pdbi-on-hst} summarize
our molecular and continuum observations of \HHth{} superimposed over the
well-known HST image~\citep{burrows96}. This figure clearly suggests that
1) the outflow is detected only through the \twCO{}~emission, 2) the
\thCO{}~\Jtwo{} line mainly traces the rotating disk (the velocity gradient
along the major axis of the optical disk is a direct signature of
rotation), and 3) the mm continuum emission, centered on the optical dark
lane, is thermal and originates from the disk.  Fig.~\ref{fig:13co-c18o}
and~\ref{fig:12co21} present the channels maps of the four observed CO
lines. Those channel maps are centered around the \HHth{} systemic velocity
of the disk: 7.25\kms{} (cf.\ Section~\ref{sec:mod:disk}). Only channels
where signal is detected are shown, \ie{} from $-0.2$ to 13.8\kms{} for
\twCO{}~\Jtwo{} and from 4.9 to 10.3\kms{} for the other CO isotopologues.
As a reference, the optical dark line and jet directions as seen on the HST
image are sketched on those figures with orthogonal lines.

\subsection{The molecular cloud}

\thCO{}~\Jone{} channel maps (Fig.~\ref{fig:13co-c18o}.b) show emission
centered on the optical disk on top of large structures, unrelated to the
outflow and/or the disk. Those large structures are clearly seen on the
\thCO{}~\Jone{} emission in the same velocity interval (\ie{} 6 to
8.4\kms{}), both in the high spatial resolution PdBI data
(Fig.~\ref{fig:13co-c18o}.b) and in medium spatial resolution but wide
field--of--view BIMA data~\citep[Fig.~2 of][]{welch00}. However, they are
not detected on the \thCO{}~\Jtwo{} channel maps
(Fig.~\ref{fig:13co-c18o}.a). We thus identify those large structures with
the surrounding molecular cloud. As the disk is expected to be hotter than
the molecular cloud, the \thCO{}~\Jtwo{} line is mainly originating from
the disk while the \thCO{}~\Jone{} emission shows both contributions.
Moreover, the molecular cloud \twCO{} \Jtwo{} emission is expected to be
optically thick, hence hiding any emission from the disk/outflow system.
Indeed, \twCO{}~\Jtwo{} channels from 6.2 to 7\kms{} are devoid of signal
in the disk/jet directions on Fig.~\ref{fig:12co21}.  The interferometer
has probably filtered out this emission because the size scale of the
optically thick molecular cloud emission is larger than typically half the
interferometer primary beam (the well--known short--spacings problem).
Finally, the emission of \CeiO{}~\Jone{} (the most optically thin line of
this study) is very weak and peaks from 6.2 to 6.8\kms{}
(Fig.~\ref{fig:13co-c18o}.c).

\subsection{The molecular disk}

\FigDiskPV{} %
\FigHCOp{} %

Contrary to \thCO{}~\Jone{}~line, the \thCO{}~\Jtwo{}~emission shown in
Fig.~\ref{fig:13co-c18o} is clearly dominated by the circumstellar disk.
The disk is seen in the \thCO{}~\Jtwo{}~emission from \about{}5 to
10\kms{}. The disk is also detected in the same velocity range in the
\twCO{}~\Jtwo{} channel maps (cf.\ Fig.~\ref{fig:12co21}).

Fig.~\ref{fig:disk-pv}.a shows a Position-Velocity plot along the disk axis
for \twCO{}~\Jtwo{} and \thCO{}~\Jtwo{}. We overplotted 1) the direction of
star and the systemic velocity as two orthogonal blue lines and 2) the
curves of the theoretical Keplerian velocity for a 0.45~\Msun{} star. Note
that this is just an illustration: The stellar mass value and uncertainty
are determined through the detailed disk analysis described in
section~\ref{sec:diskchi2}. A clear signature of the Keplerian rotation of
the disk is visible. However, some confusion exists from the parent cloud.
For \twCO{}~\Jtwo{}, this confusion is so important that the disk emission
stays undetected due to the molecular cloud between 6.2 and 7\kms{}.  For
\thCO{}~\Jtwo{}, the confusion in the disk direction is important only for
the 6.83, 6.51 and 6.19\kms{} channels and minor elsewhere. The detailed
modelling of the \HHth{} circumstellar disk described in
section~\ref{sec:diskchi2} is thus mainly based on the \thCO~\Jtwo{} data
in the velocity range devoid of confusion by the parent cloud.

With a critical density of $n_c \simeq 4 \,10^4\pccm$, the
\HCOp{}~\Jone{}~transition is expected to be thermalized in protoplanetary
disks. This transition is also easily observed in molecular clouds with a
sub-thermal excitation, and often detectable in molecular outflows.
However, the \HCOp{} \Jone{}~emission in \HHth{} only comes from the disk
as illustrated by Fig.~\ref{fig:hcop10}.

\subsection{The dust disk}
\label{sec:data:cont}

We obtained continuum maps at four different wavelengths: 3.4, 2.7, 1.35,
and 1.3~mm.  All of them show emission coincident with the \HHth{}
protostar position.  This strongly suggests we are observing thermal
emission from the dusty disk. Only the merged 1.35~mm and 1.30~mm continuum
map is shown in Fig.~\ref{fig:pdbi-on-hst}. As a first analysis, we fitted
in the $uv$ plane a bi-dimensional Gaussian function in each data set.
Results are displayed in table~\ref{tab:cont}. The continuum fluxes agree
with previous measurements made at OVRO~\citep[]{stapelfeldt99b}. A fit of
a power law, $S_\nu = S_{100} (\nu/100 \emr{GHz})^\alpha$, through the
measured fluxes yields $\alpha = 2.22 \pm 0.08$ and $S_{100} = 3.19 \pm
0.18$~mJy. Low frequency continuum emission is unresolved while a Gaussian
fit through the 1.35~mm and 1.30~mm merged data set gives a size of $ 1.26
\pm 0.11 \times 0.58 \pm 0.09''$ with a major axis at PA $-50 \pm 5\degr$
(\ie{} a disk \emph{axis} at PA $40 \pm 5\degr$).  This orientation is
consistent with the orientation of the dark lane in the optical
images~\citep[disk axis at $\rm PA~32.2 \pm 1.0\deg$,][]{burrows96}.
Assuming a distance of 140~pc, the fitted linear continuum FWHM is $\sim
175 \times 80$\au{}.

\subsection{The molecular outflow}

The outflow is only detected in the \twCO{}~\Jtwo{} emission. However, as
said before, there is confusion with the disk and parent cloud emission at
velocities close to the \HHth{} systemic velocity. Only the channels at
extreme velocities ($\leq 4\kms$ and $ \geq 11\kms$) can be attributed to
the outflow without ambiguities. The integrated image from the outflow and
the disk are presented in Fig.~\ref{fig:12co21-flux}, showing that the
outflow contributes about twice as much to the total flux as the detected
disk and parent cloud.

Fig.~\ref{fig:12co21} clearly shows that the CO outflow is essentially
one-sided. While the North-East lobe is prominent, there is no emission
from South-West, except at low level (3 to 6 sigma) in the velocity range
7.8 to 8.6\kms. In this northern lobe, the \twCO{} emission delineates a
conical structure, whose apex coincides (within the resolution of these
observations) with the star location. The semi-opening angle is
$\sim$30\deg{}.  The highly-collimated jet seen by the HST is located
precisely on the axis of this conical \twCO{} outflow, in agreement with
the usual disk/jet formation paradigm.

As expected from the edge-on geometry of the disk, the outflow lies almost
perfectly in the plane of sky. This is indicated by the outflowing gas
being observed at both blue and red--shifted velocities in the same
(northern) lobe, as well as by the quite low maximal velocities (only
\about{}6\kms{}). Indeed the extreme velocities, 1 and 13.2\kms, are almost
exactly symmetric from the systemic velocity determined from
\thCO{}~(7.25\kms{}). Assuming the \twCO{} emission arises from outflowing
gas along the cone wall, with a constant velocity expansion, we can derive
an outflow velocity of $\simeq 12 \pm 2$\kms{} and an inclination of
$\simeq 0 \pm 2\degr$.

Although the molecular outflow is definitely observed as a conical
structure emanating from the \emph{inner} part of the disk, CO ``clumps''
also exist in projection along the jet axis, \eg{} $7''$ NW off the source
as illustrated in Fig.~\ref{fig:clump}.

\FigMainCOFlux{} %
\FigOutflowClump{} %

\section{Disk Model}
\label{sec:mod:disk}

\subsection{Analysis method}
\label{sec:disk:analysis}

A standard disk model~\citep{pringle81} was used to quantify the properties
of the \HHth{} circumstellar disk.  This model assumes an axi-symmetrical
geometry, local thermodynamic equilibrium and the absence of a vertical
gradient of temperature. Turbulence is phenomenologically introduced by
adding a turbulent width $\Delta v$ to the thermal width. Radial evolutions
of temperature, surface density and velocity are assumed to be power law
normalized at 100\au{}: $T=T_{100} \paren{r/100\au}^{-q}$,
$\Sigma=\Sigma_{100} \paren{r/100\au}^{-p}$ and $V=V_{100}
\paren{r/100\au}^{-v}$.  A value of 0.5 for the velocity scaling exponent
$v$ implies that the disk rotation is Keplerian. We further assume that the
density profile follows a Gaussian distribution of the height $z$ above the
disk plane with the scale height $H(r)$ being a power law of $r$
\begin{displaymath}
  n(r,z) = n(r) \exp{\cbrace{-\paren{\frac{z}{H(r)}}^2}}
  \mbox{~and~} n(r) = \frac{\Sigma(r)}{\sqrt{\pi}.H(r)}
\end{displaymath}
\begin{displaymath}
H=H_{100} \paren{r/100\au}^{h}   \mbox{~and~}  n=n_{100} \paren{r/100\au}^{-s}
\end{displaymath}
which implies $s=p+h$~\citep[hydrostatic equilibrium would further imply $h
= 1+v-q/2$, see][]{dartois03}.

The model parameters are: $D$ the distance from Earth, $V_\emr{LSR}$ the
systemic velocity, PA the plane--of--sky orientation, $R_\emr{out}$ the
disk outer radius, $\Delta v$ the turbulent line width, and the parameters
of the power laws ($\cbrace{V_{100},v}$, $\cbrace{T_{100},q}$,
$\cbrace{H_{100},h}$ plus a pair $\cbrace{\Sigma_{100},p}$ for each
molecule transition and for the dust). Since the whole model is described
by a limited number of parameters, it is possible to perform a $\chi^2$
minimization in order to derive the best-fitted values of each parameter.
The fitting is performed in the $uv$ plane to avoid any error introduced by
the imaging and deconvolution stages. The $\chi^2$ is thus defined from the
difference between the observed and predicted visibilities:
\begin{displaymath}
  \chi^2 = \sum_{u,v,V} \abs{\frac{\emr{model}(u,v,V)-\emr{observed}(u,v,V)}{\sigma_{(u,v)}}}^2
\end{displaymath}
where $\sigma_{(u,v)}$ is the noise associated to each
visibility~\citep[see][]{guilloteau98}. The minimization were done after
subtraction of the continuum visibilities from the line $uv$
table~\citep[for details, see][]{Pietu_etal2006}.

The analysis results are summed up in Tables~\ref{tab:mm-nir-comp}
and~\ref{tab:mm-dust}. The comparison between our best model and the
observations is shown in Fig.~\ref{fig:disk:mod-vs-obs} for the
\thCO{}~\Jtwo{}~line.

\subsection{\thCO{}~lines}
\label{sec:diskchi2}

\subsubsection{Fit difficulties}

\FigModvsObs{} %

The near edge-on geometry of \HHth{} makes this source a special case.
First, the model must properly sample the disk thickness to adequately
model the line emission. Second, the number of independent data points is
actually small since the emission is unresolved perpendicular to the disk
plane. This implies a significant degeneracy between parameters $T_{100},
q, \Sigma, p, \Delta v$ and the scale height $H(r)$. In fact, the
\thCO{}~\Jtwo{}~line intensity depends weakly on the temperature in the
expected temperature range for \HHth{}, \ie{} around 20 K for the linear
scales sampled~\citep{dartois03}. At any (non zero) projected velocity
$v_\emr{obs}$, the line flux will mostly come from a radius $r = 100
(V_{100} / v_\emr{obs} )^2$, and the total emitted flux will be essentially
proportional to $\Sigma(r) H(r) \delta V$, where $\delta V$ is the local
intrinsic line width.

A further difficulty is the existence of contamination by the molecular
cloud, and perhaps also by the outflow.  We avoided contamination by the
molecular cloud by ignoring the velocity range from 6.2 to 6.8\kms{} in the
determination of the $\chi^2$. Any contamination by the outflow would
separate from the disk only spatially, not in the velocity space. As the
fit is done in the $uv$ space, contamination by the outflow is unavoidable.
Although the outflow contribution is most likely small, because emission at
large velocities from the outflow region is not detected in the
\thCO{}~\Jtwo{} channel maps (see Fig.~\ref{fig:disk:mod-vs-obs}), it may
slightly bias the inclination towards lower values.

\subsubsection{Fit steps and results}

We have performed $\chi^2$ global minimizations using the following set of
parameters: PA, $i$, $R_\emr{out}$ to constrain the geometry, and
$V_\emr{lsr}$, $V_{100}$, $v$, to constrain the velocity power law, using
fixed values for the other parameters (see Table~\ref{tab:mm-nir-comp}).
The systemic velocity value of 7.25\kms{} quoted before comes from this
step.  We verified that the rotation is Keplerian on the pair of parameters
$(V_{100}, v)$. This again suggests that any remaining confusion due to the
outflow or cloud is small. We found \HHth{} to be essentially edge on, with
a best fit inclination $\approx 81\deg$.  We nevertheless fixed the
inclination to $84\deg$ in all other fits to avoid the possible inclination
bias due to the outflow contamination (see above). We verified that all
other parameters are essentially independent on this small variation of the
inclination.

The degeneracy between the remaining parameters cannot be fully removed
with the available data. We thus assumed the height law parameters to be
$H_{100} = 22$ AU, and $h = 1.25$. We then fitted the \Jone\ and \Jtwo\ 
lines of \thCO\ simultaneously. We find for the temperature law: $T = 12
\pm 1$~K , $q = 0.55 \pm 0.07$. The associated \thCO{} surface density at
100 AU is about $9\,10^{16}$~cm$^{-2}$ (with a factor two uncertainty), but
the surface density power law index $p$ is less well constrained: $p \simeq
1.5 \pm 1$.

Temperature and density power laws are less certain because 1) the
simultanesous fit is only valid in the absence of strong vertical
temperature gradient and 2) the constraint uses only a few channels from
the \Jone\ line of \thCO{} where the contamination by the parent cloud is
thought negligeable. We nevertheless stress that our assumptions have
essentially no influence on the main disk parameters, V$_{100}$, $i$, and
R$_\emr{out}$, which are well constrained by the current data.

\subsection{\HCOp{}~\Jone{}~line}

\HCOp{} emission is detected from the disk, but with a relatively low
signal--to--noise ratio.  There is \emph{no} evidence of \HCOp{} emission
either from the cloud or from the outflow. Accordingly, the \HCOp{} data
were analyzed in a very similar way to the \thCO{}~data, but all channels
were included in the analysis.  The results (\eg{} outer radius) agree
within one sigma with those obtained from \thCO{}, but the error bars are
large. In practice, only the \HCOp{} column density is constrained from
these observations. Using $p=1.5$ and $R_\emr{out} = 400$ AU, we find
$\Sigma[\HCOp] = 7.2 \pm 1.5\,10^{12}$ cm$^{-2}$ at 100~AU.

\subsection{Continuum Emission}

We have high resolution continuum observations at four different
frequencies.  As the disk dust thermal emission is at least marginally
resolved and partially optically thin, we may obtain an {\it independent
  estimate of the outer radius and the surface density}. Matching the
surface density found in this process with the one derived from the
\thCO{}~data enables to derive the abundance of \thCO{}.

We used the disk model derived from the \thCO{} data and the four continuum
measurements to constrain the dust emissivity index, $\beta$, and the
surface density $\Sigma_{100}$ assuming the emissivity law is given by
$\kappa(\nu) = 0.1 (\nu/10^{12} \emr{Hz})^\beta$ cm$^2$g$^{-1}$.  We made a
global $\chi^2$ fit using the following four parameters: $R_\emr{out}$,
$\beta$, $\Sigma_{100}$ and $p$. 
\TableDiskPropertiesComparison{} %
\TableDustProperties{} %
\clearpage{}
We assumed identical gas and dust temperature at 100~AU. The best fit is
obtained for a uniform disk ($p=0$) of small radius ($150 \pm 20$ AU) and
total mass $2.7\,10^{-3} \Msun$, but solutions with a decreasing surface
density and larger radii remain acceptable. Assuming the same outer radius
than for CO leads to $p = 1.1$, and a surface density at 100 AU of $3.6 \pm
0.3\,10^{22}$~cm$^{-2}$, which corresponds to a mass of $4.8\,10^{-3}
\Msun$. Note that the mass scales inversely with the assumed dust
temperature. In all cases, we find $\beta = 0.4 \pm 0.1$.

\section{Outflow Model}

\subsection{Model description}

While a geometrically thin disk with power law distributions of the radial
dependence of density and temperature constitute a good description of
circumstellar disks, no such paradigm exist for outflows.  Outflow models
still rely on \textit{ad hoc} parametrization, with basic ingredients often
differing from source to source.

The \twCO{} outflow of \HHth{} presents a remarkably simple structure, both
in terms of morphology and velocity distribution that we modelled as
sketched in Fig.~\ref{fig:outflow-sketch}.

{\bf Geometry --} We assume a perfect conical geometry, with an
semi-opening angle $\theta_{\rm max}$. The cone axis is inclined by an
angle $i$ to the line-of-sight.

{\bf Velocity distribution --} The velocity of the outflowing gas is
assumed to have two components at any position: a radial component, \ie{}
this component vector is pointing outward of the central source position
and an azimuthal component (to trace a possible outflow rotation), \ie{} a
vector component tangent to the circle defined by the intersection of the
outflow cone with a plane perpendicular to the jet axis.

{\bf Density distribution and Excitation --} Our model is assuming
optically thin isothermal emission: The emissivity of the gas is
proportional to the local density and the brightness distribution is
proportional to the column densities. The model is computed in arbitrary
units, and the peak of the resulting image is further scaled to the peak of
the observed maps. Our approach is thus to focus on the kinematics and
morphology of the outflow rather than on the excitation conditions. We
assumed that the density is located mainly along the cone edges with a
constant Gaussian width scale $w_0$ and a Gaussian decrease of the density
with the distance to the star along the jet axis ($H_0$ being the height
scale).

The model is computed on a sufficiently fine grid, and is further convolved
with the clean beam of the observations, in order to mimic the same angular
resolution.

\FigOutflowSketch{} %

\subsection{Best model}

Fig.~\ref{fig:outflow:mod-vs-obs} shows a comparison of the \twCO{} data
and of our ``best'' model. This comparison is composed of images of the
integrated emission and of two position--velocity diagrams (perpendicular
to and along the jet axis) and of the spectra integrated over the outflow.
Complete sets of channel maps and position--velocity diagrams are presented
in Fig.~\ref{fig:outflow:xy}, \ref{fig:outflow:xv} and~\ref{fig:outflow:vy}
(electronic version only).  While the \twCO{} emission clearly reveals the
outflow structure, it also includes a contribution from the rotating disk
and the parent cloud.  Hence, we blanked out the velocities between 4.6 and
9.8\kms{} in the data cube in order to make meaningful comparisons between
data and model. While this is clear on the position--velocity diagrams and
the integrated spectrum of Fig.~\ref{fig:outflow:mod-vs-obs}, note that the
integrated emission map shown as the top, left panel also excludes this
velocity range. Fig.~\ref{fig:outflow:para:1} and~\ref{fig:outflow:para:2}
(electronic version only) are variations of
Fig.~\ref{fig:outflow:mod-vs-obs} which illustrate how the different
parameters of the outflow model influence the observables.

We did not develop any $\chi^2$ fitting procedure. We took instead the
following steps to successively find a plausible range of parameters
representing the data. The systemic velocity was assumed to be the same for
the outflow as for the disk. The position angle was tuned to give the best
averaged horizontal symmetry of the outflow on each side of the jet on the
channel maps of Fig.~\ref{fig:outflow:xy}. The opening angle of the cone
was measured on the $vy$ position--velocity diagrams
(Fig.~\ref{fig:outflow:vy}) as the averaged slope of the minimum $y$ value
where emission is detected as a function of the $x$ plane. The magnitude of
the turbulent line width was deduced from the slope of the outer wings of
the integrated spectrum. Indeed, Fig.~\ref{fig:outflow:para:2} shows that
the absence of turbulent line width makes infinitely sharp outer wings.
Once the systemic velocity is fixed, only the inclination on the
plane--of--sky ensure the correct velocity centering of the emission
features in the $vy$ and $xv$ position--velocity diagrams. Once the cone
opening angle and the turbulent width are fixed, only the radial velocity
can ensure the correct velocity width in the $vy$ and $xv$
position--velocity diagrams. The density width and height scales were
finally tuned to reproduce the width and height of the horn in the $vy$
position--velocity diagram. Table~\ref{tab:outflow} lists the parameters of
the ``best'' model and associated possible ranges of parameters. We checked
that the vertical limitation of the outflow is mainly physical, \ie{}
\emph{not} produced by the short-spacing filtering of the interferometer.
Finally, rotation was searched for but not found (see next section).

The model depicted in Figs.~\ref{fig:outflow:mod-vs-obs}
to~\ref{fig:outflow:vy} and Table~\ref{tab:outflow} nicely reproduces the
main characteristics of the \HHth{} outflow, both in terms of velocity and
morphology.  Hence, we conclude that, to first order, the outflowing gas in
\HHth{} forms \textbf{a cone with a constant radial velocity distribution
  and no detectable rotation} (see next section).  The inclination on the
plane--of--sky (constrained to be essentially $-1\deg \pm 1\deg$) indicates
that this cone is nearly, but not exactly, perpendicular to the disk.

Nevertheless, a number of features are not reproduced by the model. First,
our fully axisymmetric model can not reproduce the various asymmetries
seen: 1) Fig.~\ref{fig:outflow:vy} illustrates that the cone opening angle
is correct for negative $\delta x$ offsets but too small for positive
$\delta x$ offsets; 2) The position--velocity diagrams and the integrated
spectra of the data clearly shows a brightness asymmetry between the
$[0.6,4.2]$ and the $[9.8,13.4]\kms{}$ velocity ranges. Second, the
brightness distribution is more centrally peaked in the data than in the
model (maybe due to heterogeneous excitation conditions).  Third, there is
the high velocity ``clump'' 1000\au{} from the star along the flow axis.
The associated emission shows a continuity both in the red and
blue--shifted velocities with the cone emission that surrounds the optical
jet (See the $v-\delta y$ diagram in Fig.~\ref{fig:clump}).  This emission
peak thus looks like a ``clump'' in the cone of outflowing gas, in contrast
with standard CO ``bullets'' which show a distinct emission peak at
velocities significantly higher than the bulk of the outflow~\citep[See
\eg][]{bachiller90}.

Furthermore, because of the low inclination, our observations may be
insensitive to the component of motion parallel to the flow axis, because
such a gas would appear at the systemic velocity and be indistinguishable
from the surrounding cloud. Since we do not know the cloud extent, we can
neither prove nor disprove that such a confusion is happening.

\TableOutflowProperties{} %
\FigOutflowBest{} %

\subsection{Rotation?}

There is no strong prescription on the dependence of the rotation velocity
with distance from the star and the jet axis. To test the possibility of
rotation, we have introduced several rotation laws in the model: 1)
constant velocity, 2) solid rotation $V_\emr{rot} \propto r$, 3) Keplerian
rotation, and 4) vortex rotation $V_\emr{rot} \propto 1/r$ (corresponding
to angular momentum conservation in an expanding thin ring that would
generate the outflow cone). In all those cases, rotation would manifest as
a tilt of the ellipse of the $(V,\delta x)$ position--velocity diagram
(second row of Fig.~\ref{fig:outflow:mod-vs-obs}, see also
Fig.~\ref{fig:outflow:para:2}, left column). We selected as upper limit to
$V_\emr{rot}$ the value that tilted the ellipse by $5\deg$. In all cases,
we can firmly rule out values of $V_\emr{rot}$ larger than 1\kms{} at
200\au{} from the jet axis. In addition, Keplerian and vortex rotation
would give rise to low level but detectable wings outside the displayed
velocity range. Such wings are not detected in the complete data set,
probably ruling out these kinds of rotation.

\section{Discussion of stellar and disk parameters}

\subsection{Stellar mass estimate, luminosity and spectral type}

Using a distance of 140\pc{} and the magnitude of the Keplerian rotation,
we deduce a mass of 0.45\Msun{} for the young star. To use this stellar
mass value in a distance independent evolutionary track diagram
$\cbrace{\log(L_*/M^2_*), \log(T_\emr{eff})}$, we also need the stellar
luminosity value. The current estimates for \HHth{} are uncertain because
most of the stellar light is intercepted by the edge-on disk.
\citet{burrows96} derived this stellar luminosity from scattered light
emission at the disk surface, under the assumption of thermal equilibrium,
and found (using our definition of $H(r)$):
\begin{displaymath}
  L = 1.0 \Lsun (H_{100}/22 AU)^{12} (M/0.67 \Msun)^6.
\end{displaymath}
Using $H_{100} = 22$ and $M=0.45 \Msun$, the corresponding luminosity is $L
= 0.1 \Lsun$, with at least a factor 2 uncertainty due to the assumed scale
height. Based on the analysis of the spectro--energy distribution (SED),
\citet{kenyon98} have estimated the stellar luminosity to be $\sim
0.2\Lsun$. More recently, \citet{cotera01} have estimated the luminosity to
be between 0.2 and $0.9\Lsun$.

Assuming $L_*=0.2\Lsun$ as the more probable value and using
\citet{baraffe98} tracks \citep[see Fig.~3 of][]{simon00}, we find that the
spectral type seems to be M3 with an age around 4-5 Million years. Note
however that the tracks are almost parallel to the ordinate, \ie{}
$\log(L_*/M^2_*)$, in this range of mass and luminosity. A relatively small
error on the luminosity leads to an important uncertainty on the age. If we
take the other end of the luminosity range (\ie{} 0.9 $\Lsun$), we get the
same spectral type for an object which is less than 1 Million year old.
Using the~\citet{siess00} tracks, the spectral type would be M1-M2 with an
age between 4 and 1 Myrs for a luminosity equal to 0.2 and $0.9\Lsun$,
respectively~\citep[see again Fig.~3 of][]{simon00}.

\citet{appenzeller05} derive a spectral type of K7 for \HHth{}, but caution
that this results is based on a limited wavelength coverage.  From Fig.~3
of \citet{simon00}, such a spectral type is incompatible with a mass of
0.45\Msun{} for the evolutionary tracks of~\citet{baraffe98},
\citet{palla99}, and ~\citet{siess00} which fit all other sources. A K7
spectral type is compatible for the \citet{dantona97} tracks, but these
fail to give agreement for the other stars. The \citet{appenzeller05}
result nevertheless suggest that a spectral type of M3 is unlikely.

Adding the existence of a relatively powerful jet, it then seems reasonable
to classify \HHth{} as a young TTauri of stellar mass 0.45$\Msun$, spectral
type ranging around M1 and age around 4~Myrs.

\subsection{Disk Geometry}

Both the mm and the optical observations reveal a disk close to edge-on.
Table~\ref{tab:mm-nir-comp} quantitatively compares the results from both
wavelength ranges. Optical \& NIR observations indicate that the disk is
tilted along the line of sight with $i$ ranging from 82 to
84$\degr$~\citep[and references therein]{burrows96}. In all cases, the PA
is found to be 32$\degr$ counted anti-clockwise from North. Our results
agree well with these values.

The determination of the outer radius is more model dependent.
\citet{cotera01} and \citet{wood02} have assumed $R_\emr{out} = 200$\au{}
while \citet{watson04} have taken $R_\emr{out} = 250$~\au{}. All these are
consistent with the minimum radius derived from the 1.3\mm{} continuum
emission. However, in their best simulations (\ie{} \emph{not} the
canonical one), \citet{burrows96} have found $R_\emr{out} \simeq 440$\au{},
in excellent agreement with the value derived here from \thCO~\Jtwo{}.

\subsection{Disk Mass and Dust Properties}

Disk masses were previously derived by several authors from NIR and optical
scattered light images~\citep{burrows96,cotera01}, sometimes combined with
modelling of the SED~\citep[and references therein]{wood02}.

Our derived disk mass depends on three assumptions: the temperature law,
the surface density exponent $p$ and the dust emissivity.  In the optical
and NIR, only the upper layers of the disk surface are seen. Therefore, a
detail modelling of the surface density distribution cannot be achieved
without some \emph{a priori} assumptions even though the width of the dark
lane can be used to estimate the opacity law. \citet{burrows96}
extrapolated a total disk mass of $6\, 10^{-3}$\Msun{}, including a
correction factor of $\simeq 15$ to account for the likely existence of a
vertical temperature gradient.

\citet{wood02} derive a mass of $1.5\,10^{-3}$\Msun. To do this, they use a
SED fitting procedure and more sophisticated dust properties (\eg{} with a
grain size distribution) leading to an emissivity of $\sim 40$\emiss{} at
0.6~$\mu$m.  \citet{wood02} dust property implies $\beta =0.75$ and an
emissivity of $0.08$\emiss{} at 1.3\mm. The difference with our
determination is thus mainly due to the assumed dust properties (although
the temperature and disk size also have an influence).

The value $\beta = 0.4 - 0.5$ is at the low end of the range encountered in
proto-planetary disks, and very suggestive of grain growth
\citep[\eg{}][]{testi03,natta04}. As we have resolved the disk in its
radial direction, and since our model includes the vertical distribution of
the dust, this value of $\beta$ is not biased by the contribution of an
optically thick region \footnote{The disk is not resolved in height,
  however. If the dust was confined to a very thin layer (scale hight
  smaller than a few AU), and seen edge-on ($i> 87^\circ$), an additional
  correction would be necessary.}. However, the determination of $\beta$
depends on the assumption that the 3.4 mm flux is due to dust emission
only. A small contamination by some free-free emission from the jet would
drive $\beta$ towards lower values. Additional low frequency measurements
are needed to evaluate this problem and confirm the low $\beta$ value.
\citet{draine06} demonstrated that, provided the maximum grain size exceeds
about $3\lambda$ (i.e. 1 cm in our case), such low $\beta$ values can be
obtained for power law grains size distribution $n(a) \propto a^{-\gamma}$
with exponent $\gamma \simeq 3.3$, since the apparent $\beta$ index is
linked to the material emissivity $\beta_\epsilon$ by $\beta =
(\gamma-3)\beta_\epsilon$.

\subsection{Molecular Abundances and Disk Temperature}

Matching all the surface density measurements indicate a \thCO{} abundance
of $\simeq 2\,10^{-6}$, and an \HCOp{} abundance of $\simeq 2\,10^{-10}$,
with likely variations by a factor $\simeq 2$ between 100 and 400 AU, since
the column density law exponents do not necessarily match. An additional
uncertainty of a factor 2 should be added due to the uncertainty in the
dust emissivity at 1.3 mm.

The temperature of 12~K derived from \thCO{} data is significantly lower
than the value of 34~K derived by \citet{burrows96}. This is expected since
the optically thin \thCO{} emission mostly traces the disk plane, while the
scattered light traces the disk atmosphere. A decreasing temperature
towards the disk plane is expected in a disk which is optically thick to
the stellar radiation, but thin to its own re-emission. A rather surprising
fact is that, despite this low temperature, the \thCO{} abundance suggests
a low depletion of CO compared to the surrounding Taurus molecular
cloud~\citep{cernicharo87}, in contrast with what is found in most
disks~\citep[\eg{}][for DM Tau]{dartois03}.

\section{Discussion of outflow parameters}

\subsection{Outflow mass and energetics}

In the northern lobe of the outflow, the total \twCO~\Jtwo{} line flux is
15.1~\Jykms{} (using only velocities in the $[0.6,4.2]$ and
$[9.8,13.4]~\kms{}$ range and integrating spatially in the area defined in
Fig.~\ref{fig:12co21-flux}). The \thCO~\Jtwo{} line flux from the same
region and velocity range is $\approx 0.2 \pm 0.05 $\Jykms{}. This implies
a \twCO{}/\thCO{} ratio of \about{}80. Thus the \twCO{} emission is
essentially optically thin. Further assuming it is at LTE allows us to
derive the outflowing mass. A minimum flow mass of $2\,10^{-5}$\Msun{} is
found for temperatures in the range from 15 to 25\K{}, using a CO abundance
of $10^{-4}$. The outflow mass scales approximately as the temperature for
larger temperatures. The volume density in the outflow can be derived from
its total mass and an estimate of the volume it fills. Using the conical
geometry described in the previous sections and a cone thickness of 45 AU
($0.3''$) we obtain a mean density $n(\HH{}) \simeq 8\,10^{4}\pccm$, high
enough to validate the LTE assumption.

The \HHth{} CO outflow presents an obvious, extremely strong asymmetry,
since almost no signal is detected in its southern lobe. We estimate here
the \twCO~\Jtwo{} line flux to be less than about 1.5\Jykms{} (over the
same velocity range as that used in the northern lobe and in a symmetric
region). This implies that this side of the flow is more than 10 times less
massive than the northern one. Note that, while the southern optical jet
brightness is also weaker than its northern counterpart, the density ratio
has been estimated to be only $\sim 2$~\citep{bacciotti99}.

The physical parameters of the northern jet and outflow estimated in a
region of $\sim 6''$ or $\sim 800\au$ from the central source
(corresponding to the size of the maps presented in this paper) are listed
in Table~\ref{tab:jet-flow}. The values for the outflow were derived from
the observations presented in this paper. Those for the jet have been
estimated from the literature. The mean jet density was estimated to be
around $10^4\pccm$ by~\citet{bacciotti99}. Using a jet radius of
$0.09''$~\citep{bacciotti99} and a cylindrical geometry\footnote{The
  opening angle is only $2\deg$ beyond 100\au{}, so the cylindrical
  approximation is appropriate}, the jet mass up to 800\au{} is about
$2\,10^{-8}$\Msun.  \citet{burrows96} indicates that the jet knot velocity
range between $\sim$100 and 300\kms{}, so we used a typical jet velocity of
200\kms.

\subsection{Formation of the outflow -- entrainment?}

The origin of the molecular outflows is a pending problem in star formation
theories. On large scales ($\gg 1000\au$), there is strong evidence that
the observed outflows consist in ambient molecular gas that has been put
into motion by large bow shocks propagating down the underlying
protostellar jets~\citep[\eg{}][]{gueth99}.  In this classical ``prompt''
entrainment scenario, the jet transfers momentum to the outflow via the
bow-shock, and it is usually assumed that the momentum flux is conserved in
this process~\citep[\eg{}][]{masson92}. Interestingly, this is what is
indicated by our first order estimation (see table~\ref{tab:jet-flow}).

Viscous entrainment along the jet edges naturally results in conical shape
for the outflow \citep{stahler94}. However, the predicted velocities are
essentially parallel to the flow axis, while our observations are
essentially sensitive to the radial motions because the flow axis lies in
the plane of the sky. The comparison is thus impossible in practice. The
insensitivity of our observations to the gas flowing near the direction of
the jet axis also implies that the outflow momentum flux we derive above is
only a lower limit.

If the \HHth{} outflow does indeed consist in accelerated ambient gas, the
lack of CO emission in the southern lobe would point towards different
properties of the interstellar medium in this direction, that would
strongly alter the outflow formation process. In this context, it may be
worth mentioning two similarities between \HHth{} and
L\,1157~\citep{gueth96}. While the sampled scales are different (several
10$^4$\au{} for L\,1157, only a few hundreds\au{} in the \HHth{}), both
outflows exhibit 1) a more or less pronounced asymmetry between their lobes
and 2) a conical geometry and a radial velocity near the launching region.

\TableJetOutflow{} %

\subsection{Formation of the outflow -- disk wind?}

One of the most striking results revealed by our data is the fact that the
outflowing molecular gas is continuously collimated down to the very close
vicinity of the star, at spatial scales that are {\em smaller than the disk
  size} (see Fig.~\ref{fig:pdbi-on-hst}). This is somewhat difficult to
reconcile with the propagation of a large bow-shock (see previous section).
But this is naturally explained if the observed CO emission arises from
material that has been directly launched from the disk and evolves
ballistically in the first few 100\au{} from the star. This hypothesis is
also consistent with a conical flow structure, with constant radial
velocity that is observed.  Recollimation may occur at larger distances
($>1000\au$), \ie{} outside the field--of--view of our observations.

If the outflow originates from a disk wind, we can obtain a constraint on
the launch radius from the upper limit on the rotation velocity of the
outflowing gas. Indeed, \citet{anderson03} derived a general relation
between the poloidal and toroidal velocity components of cold
magnetocentrifugal wind at large distance of the star and the rotation rate
of the launching surface, independent of the uncertain launching
conditions. This relation relies on the following facts: 1) The energy and
angular momentum in the wind are extracted mostly by magnetic fields from
the rotating disk; 2) The energy extracted is the work done by the rotating
disk against the magnetic torque responsible for the angular momentum
extraction; And 3) most of the wind energy and angular momentum at
observable distances are in the measurable kinetic form. In the simple
geometry of \HHth{}, we identify the poloidal and toroidal velocities
respectively with the radial velocity $V_\emr{rad}$ and the rotation
velocity $V_\emr{rot}$ at a given radius $r$. The rotation rate of the
launching surface is the Keplerian frequency $\Omega_0 = \sqrt{G M_* /
  r_\emr{launch}^3}$.  \citet{anderson03} relation: $V_\emr{rot} =
V_\emr{rad}^2 /(2 \Omega_0 r)$, can then be transformed into
\begin{displaymath}
  r_\emr{launch} = r \cbrace{%
    2\frac{V_\emr{kep}(r) V_\emr{rot}}{V_\emr{rad}^2}}^{2/3}
\end{displaymath}
where $V_\emr{kep}(r) = \sqrt{GM_*/r}$ is the Keplerian velocity at radius
$r$. Our upper limit of $V_\emr{rot} < 1\kms$ at $r\sim$200\au{} from the
flow axis indicates $r_\emr{launch} \leq 14$\au{}.

In this scenario, the optical jet and the outflow would be associated with
different parts of the disk wind.  The jet corresponds to the inner
($<1$\au{}), densest parts of the wind, which has a smaller opening
angle~\citep[$5^\circ$ half-opening angle,][]{bacciotti99}, larger
velocities (around 300\kms), and recollimates closer to the disk
(100\au{}). The CO outflow would correspond to the outer part of the wind,
ejected at larger radii from the star (5--15\au{}), and could recollimate
at larger distances ($\gg 1000\au$). In terms of energetics, these
observations would suggest that similar momentum fluxes are extracted from
the disk at these different radii (see Table~\ref{tab:jet-flow}).

Several hypothesis could be proposed to explain the apparent discontinuity
between the jet and the outflow, \ie{} the lack of detected material at
intermediate velocities and angles. We first note that the outflow cone is
not totally empty; in fact, in the outflow modelling presented above,
adding some material in the inner part of the jet allows us to obtain an
even better fit to the data, to the price however of several new free
parameters. One possible explanation is a projection effect: The outflow
velocity field could lie more and more in the plane--of-sky (and thus
indistinguishable from the surrounding cloud) when approaching the jet
axis. The lack of significant detection could also come from temperature
gradient, that would make the \twCO~\Jtwo{} line weaker at smaller opening
angles (higher level transitions would however emit more strongly, until
the gas becomes primarily atomic).  Alternatively, this lack of observed
emission at intermediate opening angle may be genuine, indicating that the
mass loading mechanism into the disk wind is significantly more efficient
at radii $<1$\au{} (jet emission) and between $\sim$5 and 15\au{} (CO flow
emission, see radius estimate above) than in the intermediate zone $\sim$1
to 5\au{}. One could further speculate that the region of the disk at the
origin of the weak loading level is cooler, being \eg{} shadowed by an
inner puffed-up rim.

Finally, in this disk wind scenario, the strong difference between the
northern and southern CO lobes must be related to intrinsic differences in
the ejection mechanism due \eg{} to physical differences between the
northern and southern faces of the disk, or to a different magnetic
configuration. \HHth{} also presents other, second-order, asymmetries.
Optical studies indicate that the jet is currently perpendicular to the
disk axis~\citep{burrows96}; our measurements however suggest a slight
misalignment between the flow axis and the jet/disk axis, the inclinations
differing by $6 \pm 3\deg$. In the MHD wind model discussed here, this
suggests the magnetic field does not have a symmetric pattern around the
disk axis. An additional complexity is the fact that, at slightly larger
scales ($\sim 3000\au{}$), the jet is wobbling around its mean direction by
$\pm 3\deg$~\citep{burrows96}.  This is an indication of a precession (or
more generally wandering) of the ejection direction.  However, the data
presented in this paper do not probe the properties of the molecular
outflow at this scale.

\section{Conclusions}

In the \HHth{} case, a judicious use of the complementary line emission
allowed us to derive a number of very robust results: 1) The systemic
velocity of \HHth{} is 7.25\kms{}; 2) The disk outer radius is 420\au{}; 3)
The rotation vector of the disk points toward the North-Eastern jet; 4) The
disk is in Keplerian rotation; 5) The stellar mass is 0.45 \Msun; 6) A
highly asymmetric molecular outflow originates from the inner parts of the
disk; 7) The outflow material is mainly located on the thin edges of a cone
with an opening angle of $30\degr$; 8) The outflow velocity is essentially
radial with a magnitude of $12\kms$; 9) No rotation of the outflow is
detected.

The total disk mass is less certain: \about{}$4\,10^{-3}$\Msun{}. This
depends on the assumed dust emissivity. Our measurement indicate a rather
low value of $\beta$, 0.4, suggestive of significant grain growth. In
addition, but with still less certainty, the disk appears cold, around 12 K
at 100 AU, suggesting a significant temperature gradient between the disk
plane and the surface layers. Despite this low temperature, there is no
measurable depletion of CO compared to the average abundance in the Taurus
cloud, in contrast with the general trend for all other circumstellar
disks~\citep[except those around HAe stars, see ][]{Pietu_etal2006}.

The direct determination of the stellar mass improves the evolutionary
status of HH 30, but this remains hampered by the uncertainties on the
total luminosity. The most plausible solution is that of a \about{}4~Myr
old object, with a total luminosity of order 0.2 $\Lsun$, but a younger and
brighter star cannot be excluded.

Finally, the origin of the molecular outflow remains unclear, with no
conclusive arguments to distinguish between an entrainment mechanism by the
optical jet or a disk wind. The one sided nature of the outflow remains an
unexplained issue in both hypothesis.

\begin{acknowledgements}
  We acknowledge the IRAM staff at Plateau de Bure and Grenoble for
  carrying out the observations. JP thanks S.~Cabrit for useful comments on
  jet and outflow models and M.~J.~Welch for providing the \thCO{}~\Jone{}
  spectra cube of the HL Tau environment. We thank both the editor
  C.~Bertout and the referee R.~Bachiller for theirs comments that helped
  to improve the paper.
\end{acknowledgements}

\bibliographystyle{aa} %
\bibliography{5814} %

\Online{} %
\FigOutflowXY{} %
\FigOutflowXV{} %
\FigOutflowVY{} %
\FigOutflowParaOne{} %
\FigOutflowParaTwo{} %

\end{document}